# Increased Covalence and V-center mediated Dark Fenton-Like Reactions in V-doped TiO2: Mechanisms of Enhanced Charge-Transfer

*Manju Kumari[1,‡], Dilip Sasmal[1,‡], Suresh Chandra Baral[1], Maneesha P[1], Poonam Singh[1], Abdelkarim Mekki[2,3], Khalil Harrabi[2,3], Somaditya Sen[1] \**

[1]Department of Physics, Indian Institute of Technology Indore, Indore, 453552, India

[2]Department of Physics, King Fahd University of Petroleum and Minerals, Dhahran 31261, Saudi Arabia

[3]Interdisciplinary Research Center for Advanced Materials, King Fahd University of Petroleum and Minerals, Dhahran 31261, Saudi Arabia

**Abstract:**

Tuning the valence state and electronic structure of catalytically active sites is crucial for improving Fenton and Fenton-like reactions, which rely on efficient activation of $H_2O_2$ molecule. Pure $TiO_2$, however, has inadequate activity towards the $H_2O_2$ activation and is often constrained by the intrinsic electronic limitations of pristine $TiO_2$. Herein, a rational approach has been demonstrated to improve the Fenton-like catalytic performance of $TiO_2$ through multivalent vanadium (V) doping. A comprehensive characterization using X-Ray Diffraction (XRD), Raman spectroscopy, UV-Vis spectroscopy, X-Ray photoelectron spectroscopy (XPS), Electron paramagnetic resonance (EPR), and Density functional theory (DFT) reveals that V incorporation substantially alters the electronic structure of $TiO_2$. The DFT results, supported by experimental data reveal that V doping enhances Ti-O covalence and introduces mid-gap states, leading to a reduced band gap and better charge transfer. XPS confirms the coexistence of multiple oxidation states of V, which serve as active centres for activating $H_2O_2$ and generating •OH radicals. As a result, V-doped $TiO_2$ exhibits significantly enhanced dark-catalytic activity in degrading the organic dye Rhodamine B (RhB). Overall, this study provides fundamental insights into multivalent-cation-induced valence state and electronic structure modulation in TiO$_2$, offering a promising strategy for designing high-performance catalysts via defect engineering for sustainable environmental remediation.

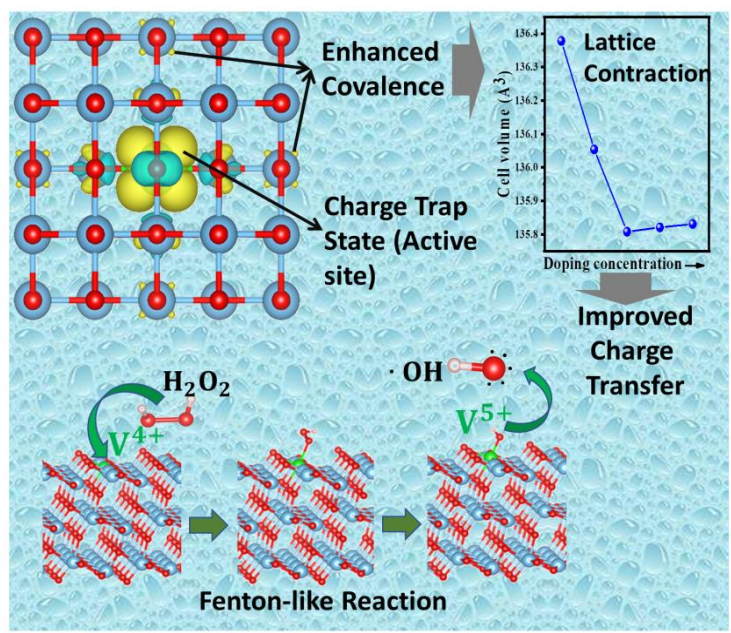

## 1. Introduction

The escalating contamination of water sources by industrial organic dyes poses a critical environmental challenge, necessitating the development of efficient and sustainable remediation technologies.[1] Among the most promising approaches, advanced oxidation processes (AOPs) utilizing heterogeneous catalysts have garnered significant attention for their ability to completely mineralize these persistent pollutants. Titanium dioxide (TiO$_2$) stands out as a benchmark material in this field, primarily due to its chemical stability, low cost, and potent photocatalytic activity, which relies on the generation of electron-hole (e-h) pairs under UV or solar illumination.[2]

While photo-catalysis is well-established, it is inherently limited by the requirement for light energy, which can be costly or inefficient in certain applications. An alternative, less-explored route is dark catalysis, which leverages chemical oxidants, such as hydrogen peroxide (H$_2$O$_2$), to drive a Fenton-like reaction for pollutant degradation in the absence of light.[3] This mechanism fundamentally requires the presence of mixed-valence states in the catalyst's constituent cations to facilitate the redox cycling necessary for H$_2$O$_2$ activation. For TiO$_2$, this implies the crucial role of intrinsic defects, specifically the presence of latent Ti$^{3+}$ sites alongside the dominant Ti$^{4+}$ species, which are essential for initiating the dark Fenton-like process.[4] The exploration of dark catalysis using pure or defect-engineered TiO$_2$ remains a significant, yet largely unaddressed, area of research.

To systematically investigate and enhance this dark catalytic potential, we propose the intentional introduction of a variable-valence dopant. Vanadium is an ideal candidate, as it can readily exist in multiple oxidation states (V$^{3+}$, V$^{4+}$ and V$^{5+}$) within the TiO$_2$ lattice.[5] This strategic doping is expected to achieve two primary goals: first, to directly introduce additional redox-active sites, and second, to induce structural defects that modify the electronic band structure and significantly increase the proportion of the latent Ti$^{3+}$ species. These modifications are hypothesized to enhance the overall mixed-valence state condition, thereby accelerating the rate of the dark Fenton-like catalytic processes. Furthermore, the creation of defect states in the bandgap is anticipated to alter the e-h recombination dynamics, which is a critical factor even in dark conditions as it relates to the stability and electronic structure of the active sites.

This work presents a systematic study correlating the structural and electronic properties of V-doped TiO$_2$ with its performance in dark catalytic degradation. By intentionally choosing V to instigate a variable valence state condition, we provide a comprehensive understanding of the underlying mechanism. Crucially, our experimental observations are substantiated by first-principles Density Functional Theory (DFT) calculations, which justify the observed structure-property relationships and elucidate the role of V doping and mixed-valence states in promoting highly efficient dark Fenton-like catalysis.

## 2. Materials and Methods
### 2.1. Synthesis of catalysts:

Vanadium substituted titanium dioxide (TiO$_2$), Ti$_{(1-x)}$V$_x$O$_2$ (henceforth referred to as TV) samples, were synthesized employing a sol-gel technique, with x=0, x=0.0312, x = 0.0468, x = 0.0625, and x = 0.125 and were named TV0, TV1, TV2, TV3, and TV4, respectively. The Ti-precursor was Dihydroxy-bis titanium (TALH) in a 50% aqueous solution (purity 99.9%, Alfa Aesar), while the V-precursor was V$_2$O$_5$ powder (purity 99.9%, Alfa Aesar). To initiate the synthesis, TALH was dissolved in deionized water (DIW), while stoichiometric quantities of V$_2$O$_5$ were dissolved separately in an aqueous solution of HNO$_3$. The precursor solution was then combined and stirred for 2 hours, ensuring a homogeneous mixture. After that, 2 grams of citric acid and 5 mL of ethylene glycol were mixed to each precursor solution in the interval of 30 minutes and stirred continuously for an hour to get homogeneously mixed solution. The homogeneously mixed solution was heated at 80°C to form the gel-former solution. The mixing continued during this heating process. Monomers were formed from the gel-former molecules to which the precursor ions got attached to. Upon further heating, the monomers polymerized to form a polymeric solution resulting in gel formation with dehydration. The attached precursor ions in the polymeric chains confirm the arrest of homogeneity of the precursor ions. Upon further dehydration of the gels resulted in the breakdown of polymeric chains, providing the necessary energy for the formation of chemical bonds between the uniformly distributed ions. This enables the formation of exceptional chemically homogeneous samples of the desired formula. The gels were thus burnt in ambience to yield black powders. During this combustion process, nitrogenous and carbonaceous oxides were released. However, traces of nitrogenous and carbonaceous components persisted, necessitating further treatment. These powders were subsequently heated at 450°C (723K) for 2 hours, effectively removing the residual impurities, and yielding white powders for the pure and darker yellowish-brown powders for the V-doped $TiO_2$ samples. This synthesis process is economical, straightforward, requires simple processing equipment, and therefore, is a practical and efficient route.

### 2.2. Characterization of catalysts:

X-ray diffraction (XRD) was performed employing a Bruker D2 Phaser diffractometer in the 2θ range of 20° to 80°, using a Cu Kα radiation of wavelength (λ) of 1.54184Å. A 30 kV operating voltage producing a current of 10 mA was employed. Further, Analysis of the XRD data were done by performing Rietveld refinement using Fullprof software. The powders' surface morphology, particle shape and size were studied using a Field Emission Scanning Electron Microscope (FESEM) [ZEISS Gemini FESEM]. The Specific Surface Area (SSA) and surface porosity were studied from the adsorption-desorption isotherms of the materials, using nitrogen as adsorbate gas by employing Quanta chrome, Autosorb iQ2, BET Surface Area & Pore Volume Analyzer. A HORIBA Scientific LabRAM HR evolution spectrometer equipped with a 633-nanometer laser light was used to study the lattice dynamics of the systems and the corresponding changes upon V doping. Diffuse Reflectance Spectroscopy (DRS) was performed at room temperature to study the optical bandgap and disorder in the lattice. A Shimadzu (UV-Vis 2600 i) spectroscope was used to measure the optical reflectance spectra of the nanoparticles in the range of 400 nm to 900 nm. The oxidation states of elements were investigated using Thermo-Scientific Escalab 250 Xi X-ray photoelectron spectroscopy

(XPS) spectrometer. Al-Kα monochromatic X-ray source with energy 1486 eV was used in this study. A flood gun was used during the experiment to counteract the effects of sample charging due to the loss of electrons. Ar$^+$ ion bombardment on the samples, for a period of 20s at an accelerating potential of 2.0 kV, was used to remove surface contamination. A survey scan was collected with pass energy of 100 eV to confirm the existence of peaks related to the elements in the samples. The high-resolution XPS spectra of O1s, Ti2p and V2p, were obtained with 20 eV pass energy. The base pressure was approximately $1 \times 10^{-7}$ mbar during the ion bombardment while the original pressure in the analysis chamber was $1 \times 10^{-10}$ mbar. All spectra were calibrated using the adventitious C 1s with a binding energy of 284.8 eV. XPSpeak4.1 software was employed to curve fit the spectra and deduce the various parameters such as peaks binding energy, full width at half maximum (FWHM) of the analysed peaks and percentages of various species after curve fitting. The EPR measurements were performed using JES-X320 with sensitivity 5 x 10$^9$ spins/0.1 mT in X-band: 8.75-9.65 GHz having magnetic field range 1300 mT.

### 2.3. Characterization of catalytic activity:

Catalytic properties of the samples on Rhodamine B (RhB) dye were studied. A 10-ppm dye concentration solution was prepared by adding 10mg RhB dye to 1L DIW in a beaker which was thereafter stirred for 2hrs. The photo-catalytic activity of the TV samples was examined by adding 10 mg of the TV samples to the 10mL solution of the dye solution. The beakers were kept in the dark for another 1hr until adsorption equilibrium of the dye molecule on the catalyst's surface is reached. 2ml of $H_2O_2$ was added and initial and final absorption spectra of the dye solutions were recorded using a UV-Visible spectrophotometer in the interval of 30 minutes. There were six beakers used: four of them contained vanadium-substituted samples, one contained pure $TiO_2$, and no catalyst was added to the remaining one in order to observe the effect of $H_2O_2$ solely. Data had been recorded for a total period of 150 minutes.

### 2.4. Computational Methodology:

Density functional theory (DFT) calculations have been performed using Vienna ab-initio Simulation Package (VASP) simulation package.[6] Projected Augmented Wave (PAW),[7] which is a generalization of the pseudopotential and linear augmented-plane wave method, has been employed to describe the interaction between frozen ion core and valency electrons with a plane wave basis set with a kinetic energy cutoff of 520 eV. Perdew-Burke-Ernzerhof (PBE) is a functional in the framework of generalized gradient approximation (GGA) and has been used for exchange-correlation (XC) interaction.[8,9] To explore bulk electronic structure and the effect of doping, a 2×2×1 supercell of anatase TiO2 (space group: I4$_1$/amd) containing 48 atoms has been used for pristine TiO2, and one Ti atom has been replaced by V for 6.25 atomic % V-doped TiO2. Monkhorst-Pack k-grid of 4×4×3 with a k-mesh resolved value of 0.03 2π/Å was employed,[10] and all the structures were fully relaxed until the Hellmann-Feynman forces on each atom were less than $10^{-3}$ eV/Å and the convergence criterion for the electronic structure iteration was set to be $10^{-6}$ eV.[11]

A slab model of 216 atoms exposing (101) crystal plane and gamma point calculation is reasonable for exploring catalytic activity. To minimize the interactions between periodic images, a vacuum space of 20 Å was set along [101], and Grimme's method (DFT+D3) was employed to consider van der Waals (vdW) interactions during surface relaxation and adsorption process.[12,13] Dipole correction along [101] has been considered to account for the artificial electric field arising from the polar surface.[14] All the slab calculations were converged until the Hellmann-Feynman forces were below $10^{-2}$ eV/Å, and the total energy change between electronic iterations was less than $10^{-4}$ eV. For transition metal oxides, Hubbard correction plays a crucial role as conventional Density Functional Theory (DFT) often fails to accurately describe the strong on-site electron-electron interactions, leading to significant deviations in predicting their electronic and magnetic properties. Hence, the incorporation of self-interaction correction to d orbital of Ti and V via Dudarev's rotationally invariant DFT + (U-J) formalism remained important throughout the project.[15] The adopted U, J values for Ti and V are 3.925 eV, 0.511 eV, and 3.4 eV, 0 eV, respectively. Ti (3d,4s), V(4s,3d) and O (2s,2p) were configured as valence state. Post-processing of the simulated data was performed using VASPKIT,[16] and VESTA[17] was used as a visualization tool for spin density, charge density differences, structural changes, etc.

The adsorption energy ($E_{ads}$) of adsorbate was defined by the equation: $E_{ads} = E_{Total} - E_{Slab} - E_{Molecule}$, where, $E_{Total}$ is the total energy of the slab together with the adsorbate, $E_{Slab}$ is the total energy of the surface exposing slab, and $E_{Molecule}$ is the total energy of the free molecule in the gas phase.[18,19] A negative adsorption energy is an indicator of a thermodynamically favoured exothermic adsorption process.

3. Results and Discussions:
3.1. Crystal structure:

The XRD patterns of the TV samples [Figure 1(a)] revealed a good match with the JCPDS card #782486, confirming the tetragonal anatase phase of $TiO_2$ (space group: $I4_1/amd$).[20] No extra peaks belonging to possible impurity phases were observed in the XRD data of all the samples, confirming the phase purity of the materials. The inset of Figure 1 (a) reveals a shift of the (101) peak towards higher 2θ values for TV1 and TV2 but finally shift towards lower values for TV3 and TV4. The peaks continuously broaden with doping, i.e. the Full Width at Half Maxima (FWHM), $\beta_{total}$ of the peaks increases. The Williamson-Hall equation: $\beta_{total}(cos(\theta)) = 0.9\lambda/D + 4\varepsilon sin(\theta)$, relates the crystallite size, D, and microstrain ($\epsilon$) to the values of $\theta$ and $\beta_{total}$ of the XRD spectra, where, $\lambda = 1.5414$ Å is the wavelength of Cu-K$_\alpha$ radiation.[21] The proper values of θ and $\beta_{total}$ of the peaks were obtained from Rietveld refinement with a Pseudo-Voigt peak shape [Figure 1(d) and supplemental material: Figure SM1]. The crystallite size of the samples seems to decrease with doping from 176.7 nm (TV0) to 85.0 nm (TV1), to 80.1 nm (TV2), to 78.4 nm (TV3) and to 76.6 nm (TV4). The lattice strain of the samples seems to increase with doping from 0.0007 (TV0) to 0.00188 (TV1), to 0.00199 (TV2), to 0.00204 (TV3) and thereafter decrease to 0.00172 (TV4) [Figure 1(b)].

Pristine $TiO_2$ is built from $Ti^{4+}$ (ionic radius ~ 0.605 Å) ions that are six-coordinated in an octahedral environment. When vanadium is introduced, it can exist in multiple oxidation states: $V^{3+}$ (~0.64 Å), $V^{4+}$ (~0.58 Å), and $V^{5+}$ (~0.54 Å), each with a slightly different ionic radius compared to $Ti^{4+}$. These size differences can locally stretch or compress the lattice. From the valence state aspect, a $V^{5+}$ state is most likely to reduce latent oxygen vacancies ($V_O$), and increase cationic vacancies or introduce oxygen interstitials ($O_i$). On the other hand, a $V^{3+}$ ion is most likely to increase $V_O$ and reduce cationic vacancies or $O_i$. Such an influx of different defects due to variations in the cationic charge and size associated with different possibilities of defects are most likely the reason for the increased strain. As a result, the crystallite size has decreased radically from about 176.7 nm to ~ 80 nm in the doped samples.

The anatase-$TiO_2$ structure is composed of edge-shared $TiO_6$ octahedron aligned along the 001 direction [Figure 1(h)]. The $TiO_6$ octahedron has two types of Ti-O bonds, two longer epical Ti-O2 bonds along the [001] crystal axis, and the other four shorter planer Ti-O1 bonds in a plane perpendicular to [001] [Figure 1(g)]. However, these four shorter Ti-O1 bonds are not perpendicular to the Ti-O2 bonds. The planer bonds seem to be decrease with doping, while the two epical bonds increase slightly for samples TV1 and TV2, then again decrease for TV3 and TV4 [Figure 1(f)]. These structural adjustments lead to an overall contraction of the lattice, as reflected in the changes in both the lattice parameters [Figure 1(e)] and the unit cell volume [Figure 1(c)]. The unit cell volume, decreases from 136.37(8) Å$^3$ (TV0) to 136.05(4) Å$^3$ (TV1), to 135.80(8) Å$^3$ (TV2), and thereafter nominally increase to 135.82(1) Å$^3$ (TV3) and to 135.83(1) Å$^3$ (TV4). This doping-induced lattice contraction can strengthen the metal-oxygen covalence, which in turn may improve the material's charge-transfer properties.

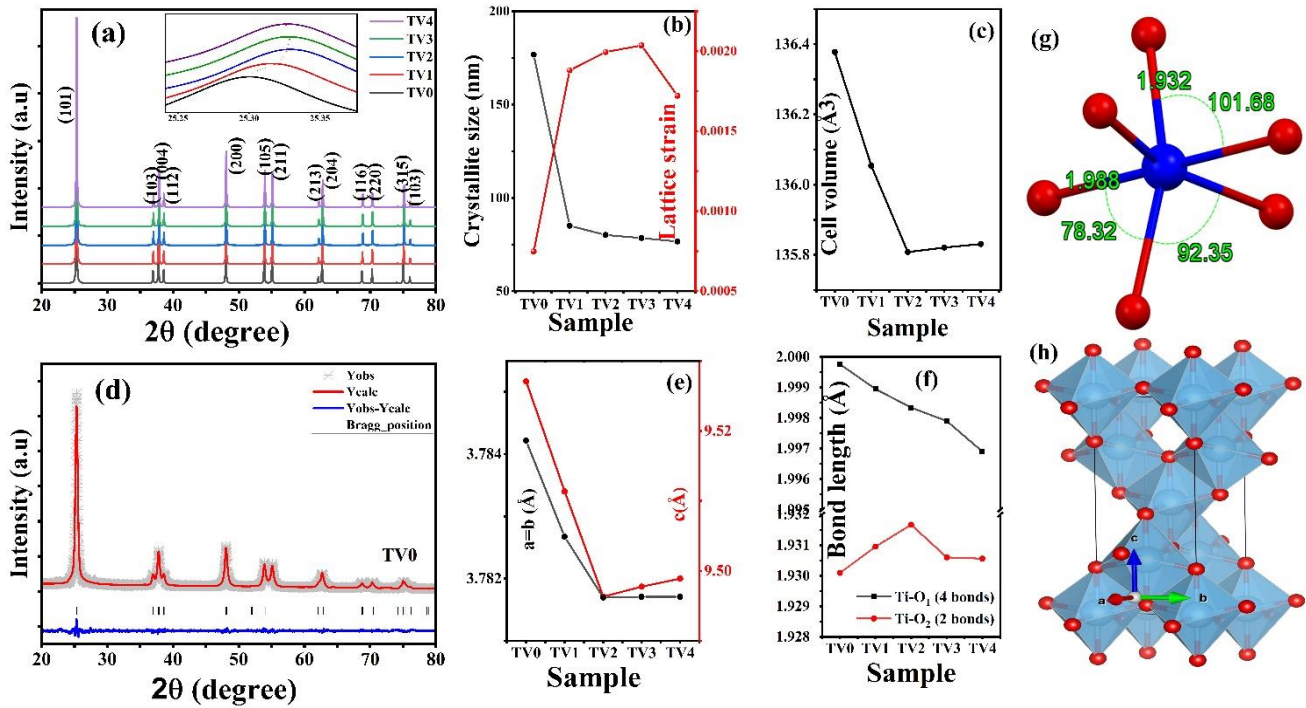

Figure 1: (a) powder XRD patterns at room temperature (~ 300 K) (inset figure shows the shifting in peak position for (101) plane). (b) Variation in crystallite size and lattice strain. (c) unit cell volume with doping concentration (d) Reitveld refined pattern of pristine TiO$_2$ (black bars indicate Bragg peaks). (e) lattice parameters (f) bond lengths. (g) TiO6 octahedron representing the two types of bond length (epical and planer). (h) Building block of anatase-TiO$_2$ i.e. edge sheared TiO6 octahedron aligned along [001] crystal axis.

According to group theory six Raman-active phonon modes are allowed for the anatase-TiO$_2$: $\Gamma_{Raman} = A_{1g} + 2B_{1g} + 3E_g$.[22] These six phonon modes are observed in these samples, as well [Figure 2(a)]. For the TV0 sample, these modes were observed at 143 cm-1($E_g$), 195 cm$^{-1}$ ($E_g$), 396 cm$^{-1}$ ($B_{1g}$), 514 cm$^{-1}$($A_{1g}$), 519 cm$^{-1}$ ($B_{1g}$) and 637 cm$^{-1}$($E_g$). All the $E_g$ modes are vibrations along the 'a' axis. The most prominent and important $E_g$ mode at 144 cm$^{-1}$ represents less energetic vibrations of the Ti and O atoms. The $E_g$ mode at

195 cm$^{-1}$ is extremely weak in amplitude but is due to more energetic vibrations involving larger displacements of the Ti and O atoms. Note that the A$_{1g}$ mode at 514 cm$^{-1}$ and B$_{1g}$ mode at 519 cm$^{-1}$ are close enough to be distinguished. The most energetic E$_g$ mode at 637 cm$^{-1}$ is somewhat more intense than the E$_g$ (195 cm$^{-1}$) and is highly sensitive to changes in the lattice. The B$_{1g}$ phonon is due to O-Ti-O symmetric bending vibrations, while the A$_{1g}$ corresponds to O-Ti-O anti-symmetric bending vibrations.[20]

The intensities of all the phonons reduce drastically after doping, implying a degradation of the crystallinity of the V-doped TiO$_2$ samples [Figure 2(b)]. The E$_g$ mode at 143 cm$^{-1}$ exhibits a clear blue-shift upon V-doping. Moreover, the FWHM of the phonon modes increases, implying a shortening of the long-range order of the crystalline phase [Figure 2(c)]. The observed blue-shift of the phonon modes indicating an increase in their vibrational energy and can be attributed to enhanced metal-oxygen covalence, which strengthens the corresponding bonds. The phonon frequency, $\omega$, is dependent on the bond strength ($k$) and the effective mass ($\mu$) of the vibrating atoms, where it can be modelled as $\omega = \sqrt{k/\mu}$.[23] The relative atomic masses of Ti (47.867) and V (50.942) are quite similar. Therefore, changes in the reduced mass ($\mu$) are unlikely to affect the phonon frequency in any significant way. Instead, the observed modification in phonon behaviour is primarily governed by bond strengthening arising from enhanced metal-oxygen covalence, which directly influences the lattice vibrations.

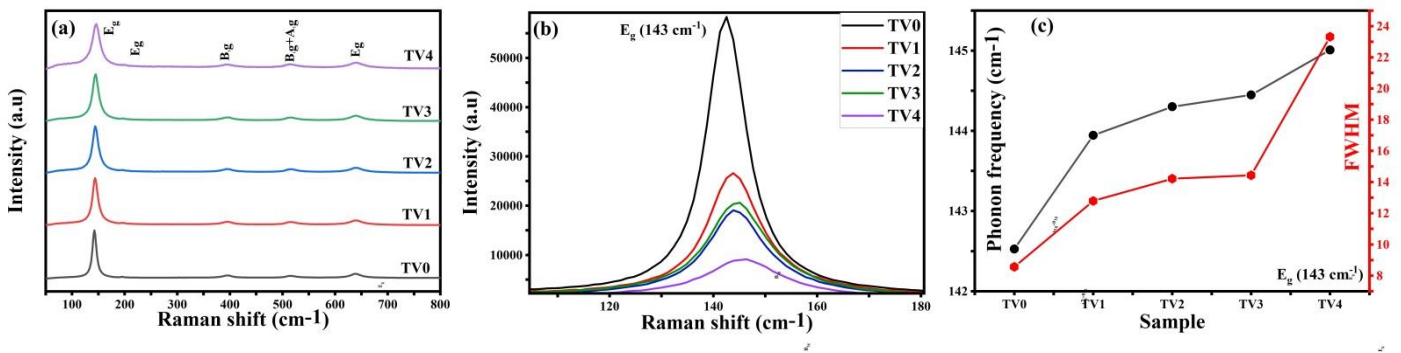

Figure 2: (a) Ambient Raman spectra of pristine and V-doped samples. (b) Variation in intensity (as-recorded data) of highly intense E$_g$ mode with doping concentration (c) Evolution of phonon frequencies and line-widths with doping concentration at room temperature (~ 300 K).

### 3.2. Surface morphology (FE-SEM) and specific surface area (BET):

To investigate whether the internal changes in the long-range order have affected the macroscopic particle size, a FESEM study of the morphology of the samples was performed [Figure 3(a)-(e)]. An agglomerated morphology was observed for all the samples, including TV0. The surface of this agglomerated bulk seems to be composed of smaller particles of size < 15nm [Figure 3(f)]. This feature is observed for all samples. As it is now known that the crystallite size of the pure $TiO_2$ is ~180nm while that of the doped samples is ~80 nm, much larger than 15nm, hence, 15nm cannot be the particle size of these materials. Hence, the morphology must represent some features of some leftover amorphous residue of the anatase phase. Therefore, an attempt is made to understand the particle size using an indirect BET analysis.

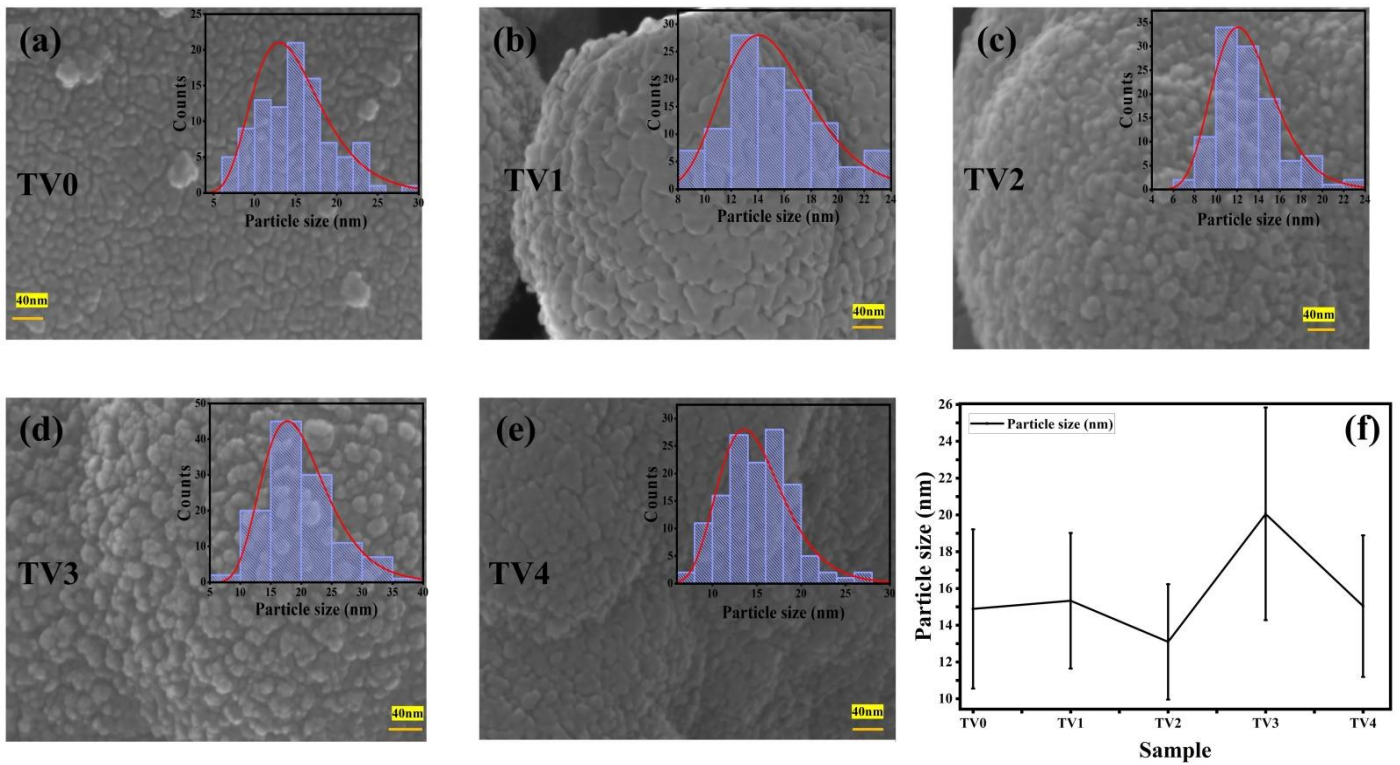

Figure 3: FE-SEM images of (a) TV0 (b) TV1 (c) TV2 (d) TV3 (e) TV4 samples (inset histograms represents the distribution of particle size as analysed by using ImageJ software and fitted by log-normal distribution). (f) Variation in average particle size with doping concentration.

To measure the specific surface area and surface porosity of the materials, N2 adsorption-desorption BET (Brunner Emmett Teller) isotherm [Figure 4(a)-(e)] measurements was performed. The samples exhibited Type IV hysteresis loops (following IUPAC classification) having irreversible adsorption/desorption processes, a signature of capillary condensation in larger pores.[24] This is supported by the noticeable rise in the $N_2$ adsorption/desorption measurements when the relative pressure gets closer to one. The "BET constant", 'C', is related to the adsorption energy in the first layer and hence, the strength of adsorbent-adsorbate interaction.[25] The value of C increases from 10 in TV0 to 70 in TV1, 98 in TV2, 100 in TV3 and

103 in TV4. For the TV0 sample, the value of C is < 20, which implies that the first layer's adsorption energy is extremely higher than subsequent layers. Amongst the doped samples, TV1 has C ~ 70 while the higher doped TV2, TV3 and TV4 samples have C ~ 100. A value of 20<C<100 represents a regime where a moderately strong adsorbent-adsorbate interaction is expected. Hence, in the V-doped samples the BET analysis is valid. A value of ~100-200 is moderately porous and is considered as a normal material, while C >200 implies significant porosity and high surface area. Therefore, these samples are moderately porous and do not have very high surface area exposed. The less porous nature implies a compact material, meaning the nanoparticles are extremely agglomerated and thereby stick to each other in maximum capacity, which may be a result of surface charge dependent attraction. Such an agglomerated nature is also substantiated by FESEM studies, where the particles seem to be united to form continuous matter.

The monolayer capacity ($V_m$) is a surface property of the adsorbent, revealing the surface area and the binding site strength. The $V_m$ of the samples was relatively lower for TV0 (2.26 mmol/g), which increased to a higher value in TV1 (21.54 mmol/g), TV2 (21.08 mmol/g), TV3 (18.33 mmol/g), and again increases TV4 (19.34 mmol/g) [Figure 4(f)]. A higher $V_m$ implies higher adsorption efficiency, which is desired in catalysis, water purification, gas separation, etc. Hence, introduction of V in $TiO_2$ probably enhances the effective surface area and the capability of binding adsorbates to the surface binding sites. Therefore, either the V itself or the structural changes of the modified lattice are responsible for the probable better applicability of the doped samples. The specific surface area (SSA) is an important parameter that needs to be understood from the above observations.

The specific surface area (SSA) increases drastically from 10.64 $m^2$/g in TV0 to 106.5 $m^2$/g in TV1 and 169.9 $m^2$/g in TV2. Thereafter, it is reduced to 85.89 $m^2$/g in TV3 and 97.78 $m^2$/g in TV4. Again, there is a drastic difference in the SSA between pure and V-containing materials. If the particle size is smaller, the surface-to-volume ratio increases and a high surface porosity is obtained. The pore volume and pore radius were analysed using the Barrett-Joyner-Halenda (BJH) model [supplemental material: Figure SM2 and Table SM2].[26] The pore radius was calculated to be TV0 (15.31 Å), which increases for TV1 (24.64 Å), TV2 (24.58 Å), TV3 (24.66 Å) and TV4 (24.57 Å), with pore volume being TV0 (0.022 cc/g), which increased for TV1 (0.122 cc/g), TV2 (0.131 cc/g), TV3 (0.104 cc/g) and TV4 (0.128 cc/g).

These results suggest that V doping enhances the adsorption capability of the material by increasing the pore volume and, in turn, the specific surface area. This ultimately leads to a higher monolayer adsorption capacity. An increase in the SSA in these samples implies a reduction of the size of the nanoparticles with V-doping. It has been reported that for perfect, non-porous spherical particles, it is possible to theoretically estimate the particle size, $D$, from the SSA by an inverse relationship $D \propto 1/\text{SSA}$.[27] However, for imperfect materials, variations can be expected from the ideal relationship. Nevertheless, according to this theoretical assessment, the particle size should be drastically reduced due to increase in SSA for the V-doped samples w.r.t. the pure TV0 sample. This is what has been observed for the crystallite size of the particles. Hence, it may be possible that the long-range order of the V-doped sample is reduced due to strain and other factors, thereby reducing the particle size.

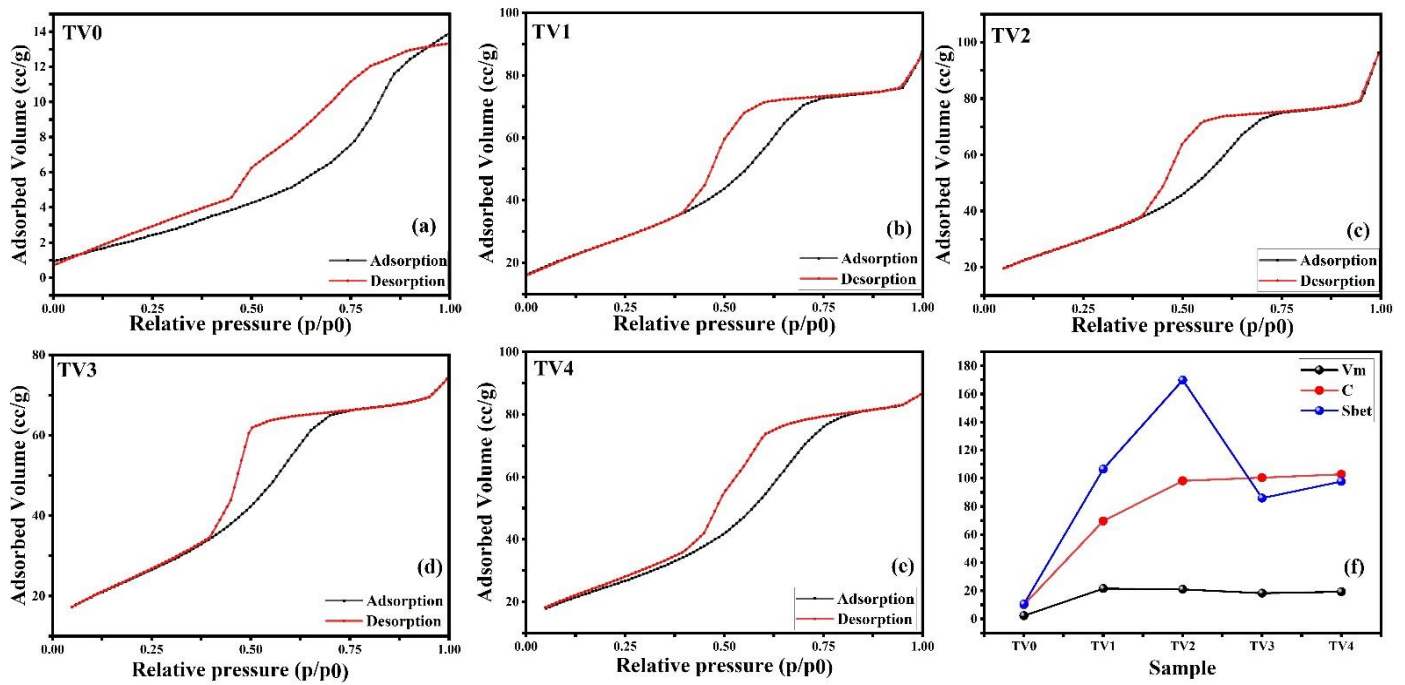

Figure 4: $N_2$ adsorption-desorption isotherm for (a) TV0 (b) TV1 (c) TV2 (d) TV3 (e) TV4. (f) Variation in monolayer capacity ($V_m$), BET constant (C), and specific surface area ($S_{BET}$) with doping concentration.

### 3.3. Electronic structure and valence state analysis:

The band-gap, Eg, of the materials were estimated from the Tauc plots [Figure 5(a)]. The Tauc relation was used to evaluate Eg: $(\alpha h\nu)^{1/n} = B(h\nu - E_g)$, where, $\nu$ is the frequency of the incident light, while B is the absorption constant that depends on the transition probability and $\alpha$ is the absorption coefficient.[28] $\alpha$ was estimated from the Kubelka-Munk function (F(R)): $\alpha \sim F(R) = \frac{(1-R_\infty)^2}{2R_\infty}$, where R is the reflectance from the sample.[28] Index, n, in the Tauc relation, characterizes the nature of the electronic transition. For a direct allowed transition, n=1/2, while for an allowed indirect transition n=2.[28] Assuming a direct band-gap, $E_g$ was found to nominally increase initially from TV0 (3.17 eV) to TV1 (3.22 eV) and thereafter continuously reduce for TV2 (3.12 eV), TV3 (2.91 eV), and TV4 (2.59 eV). Note that the $E_g$ does not significantly change for TV0, TV1, and TV2, but reduces considerably for TV3 and TV4 [Figure 5(b)].

An oxygen-vacancy ($O_V$) is an inevitable point defect in transition metal oxides.[29] In V-containing samples it is common to find a mixed oxidation state ($V^{3+}/V^{4+}/V^{5+}$) of the V-ions to be discussed later. Such a variation in the valence state can be held responsible for modulating oxygen-related defects. Oxygen defects along with the presence of multiple states of V are responsible for creating the energy state inside the band-gap or even within the valence and conduction bands. These mid-gap states lower the band gap and move the absorption edge toward higher wavelengths, bringing it into the visible region. Because these mid-gap states are associated with lattice disorder, quantified by Urbach energy ($E_U$), defined by the relation: $\alpha = \alpha_0 e^{-(h\nu/E_U)}$.[30] The Urbach energy increases continuously with the doping percentage, ranging from 190 meV to 374 meV [Figure 5(b)].

DFT electronic structure calculations are essential for identifying and confirming the mid-gap states introduced by doping, providing a direct link between the reduced band gap and increased Urbach energy observed in UV-Vis spectroscopy. The lattice parameters of the relaxed pristine anatase-$TiO_2$ (a = b = 3.892 Å, c = 9.698 Å) show slight deviations from experimental values, as they correspond to a defect-free, single-crystalline system. Consequently, the planar (1.989 Å) and apical (2.014 Å) Ti–O bond lengths also differ from those observed experimentally. The electronic band structure and partial density of states (PDOS) confirm the contributions of Ti 3d and O 2p orbitals to both the valence and conduction bands, indicating the covalent nature of the bonding [Figure 5(d)]. However, the conduction band is primarily dominated by Ti 3d states, while the valence band is mainly contributed by O 2p states. As expected, the DFT formalism underestimates the band gap, leading to a noticeable deviation from experimental values. The calculated band gap for the pristine system is 2.33 eV, which aligns well with previous reports.[31]

The valence band of V-doped TiO2 comprises O 2p states along with a filled band predominantly contributed by V 3d states located above O(2p) and just below the Fermi level. Vanadium ($V^{+4}$) has five valence electrons, of which four participate in chemical bonding, replacing Ti. The remaining electron remains localized at the V(3d) site, as confirmed by the double dumbbell-shaped spin density plot [Figure 5(c)]. This additional electron contributes to the V 3d states. The conduction band is predominantly composed of Ti 3d, with the V 3d levels situated below those of Ti 3d [Supplemental Material: Figure SM3(b)]. Differences in the properties of the dopant (V) compared to the host element (Ti) lead to modifications in both the structural and electronic properties of the material. Since V is more electronegative than Ti, the increased electronegativity of the dopant enhances the covalence between V and neighbouring O atoms. This increased covalence results in a reduced band gap, which explains why the V 3d states appear below the Ti 3d states. Note that, one of the most important consequences of V doping in $TiO_2$ is the enhancement of the Ti-O covalence, as confirmed by the positive spin density around Ti atoms in the vicinity of the dopant, arising from the increased local charge density [Figure 5(c)]. The band gap of the doped system decreases significantly to 1.7 eV [Figure 5(e)], which aligns well with experimental findings.

The formation of an oxygen vacancy reduces the valence state of neighbouring $Ti^{4+}$ to $Ti^{3+}$ and $V^{4+}$ to $V^{3+}$, leading to a decrease in the covalent character of the $TiO_6$ and $VO_6$ octahedron. This reduction in covalence, shift the conduction band to higher energies, leading to an increase in the band-gap. Simultaneously, a defect state-primarily dominated by Ti 3d orbitals emerges within the band gap due to the presence of non-bonding electrons on Ti atoms adjacent to the oxygen vacancy [Figure 5(f)].

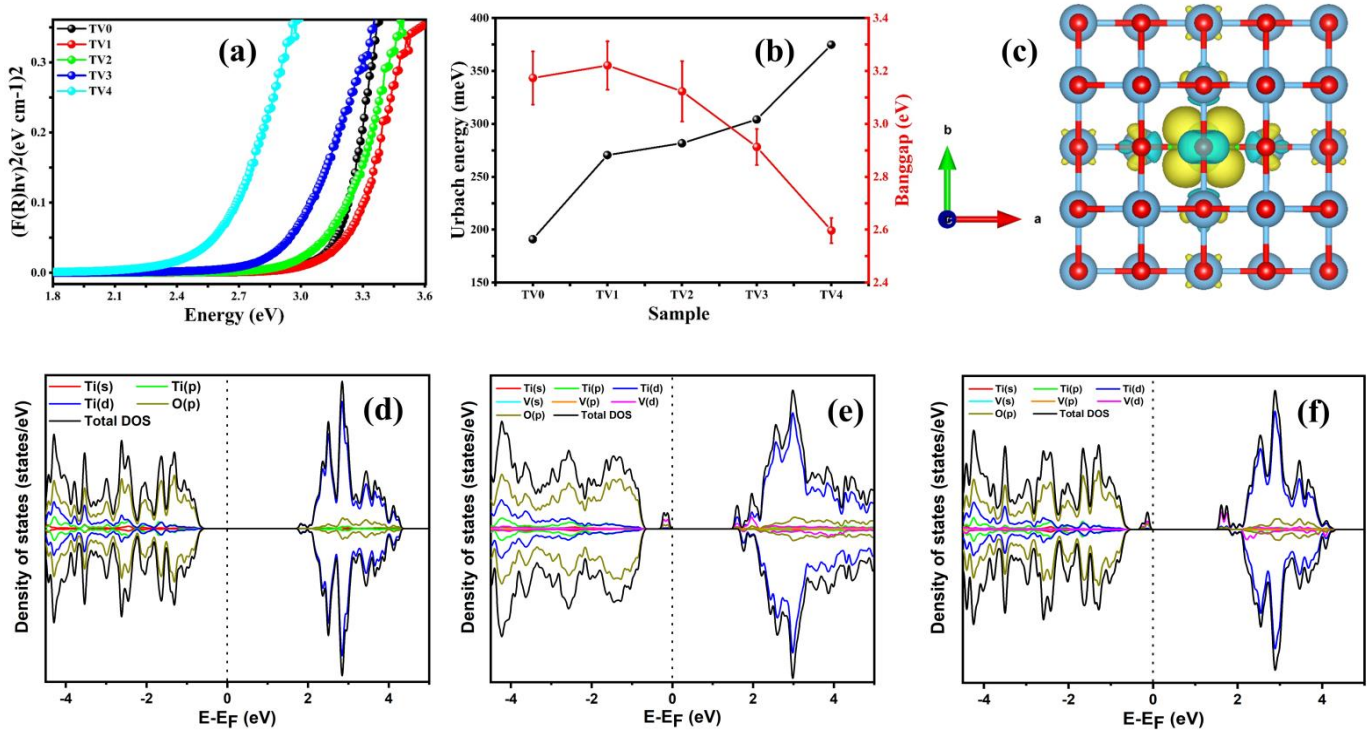

Figure 5: (a) Tauc plot for the calculation of direct band-gap. (b) Variation in band-gap ($E_g$) and Urbach energy ($E_U$) with doping concentration. (c) Spin-density plot of V-doped $TiO_2$ showing the localized unpaired electron in the characteristic dumbbell-shaped V-3d orbital, along with enhanced electron density around the Ti atoms linked to the V site through oxygen, indicating stronger Ti–O covalence. (Isosurface level: 0.001, isosurfaces represent in yellow and cyan colour signifies charge accumulation and depletion respectively) (d) Orbital projected spin-polarised density of states (DOS) for pristine $TiO_2$, (e) V- doped $TiO_2$, (f) V-doped $TiO_2$ with oxygen vacancy ($O_V$). (Fermi energy is set at 0 eV).

$TiO_2$ treated with $H_2O_2$ (even in the dark) can form •OH through redox reactions with $Ti^{3+}$. The electronic configuration of $Ti^{3+}$ ion is [Ar] $4s^0 3d^1$. This 3d electron is localized in the $t_{2g}$ orbital of the octahedral crystal field in anatase-$TiO_2$.[32] The unpaired electron has spin S=1/2 and shows magnetic moment ~1.73 μB . However, the chances of this isolated electron of a $Ti^{3+}$ ion to be paired with another electron from a neighbouring $Ti^{3+}$ ion is rare due to the low concentration of $Ti^{3+}$ content in the lattice. Hence, this magnetic moment with no complete spin pairing shows paramagnetic character. Although, the existence of $Ti^{3+}$ was not resolved from the XPS studies thereby, electron paramagnetic resonance (EPR) plays a vital role for confirming the presence of paramagnetic centres in the un-doped $TiO_2$. This hints at the simultaneous presence of oxygen vacancies. However, presence of such anionic and cationic defects are key contributors to enhanced electronic and catalytic/photo-catalytic performance due to their role in charge trapping, defect-mediated adsorption, and reactive oxygen species (ROS) generation. EPR spectroscopy is a powerful tool to understand the nature of the defects and has been widely studied in literature, differentiating the surface or bulk nature of the $Ti^{3+}$ ion.[33]

EPR spectroscopy was employed to identify and characterize paramagnetic defects in un-doped TV0 and doped TV4 samples. The room temperature EPR spectrum at resonance condition was used to calculate the Lande g-factor (g-value): $g = \frac{h\nu}{\mu_B B_r}$, where h is Planck's constant (6.626 × 10$^{-34}$ J·s), ν is the microwave frequency (9.5 GHz for X-band), $\mu_B$ is the Bohr magneton (9.274 × 10$^{-24}$ J·T$^{-1}$), and $B_r$ is the resonance magnetic field in Tesla [Figure 6(i)].[34,35] The g-factor indicates how the magnetic moment of an unpaired electron reacts to an applied magnetic field. In a typical oxygen-defect-free anatase-$TiO_2$ sample, the EPR signal should not appear at X-band because there are no unpaired spins in the material.[36] For pure TiO2 a g-factor of ~2.002 is reported in literature.[37] similar to these reports both V-doped and un-doped anatase-$TiO_2$ was found to have g ≈ 2.002. A free electron has a g ≈ 2.0023. Hence, the signature of an EPR signal at g ≈ 2.002 reveals the presence of a free electron in the lattice which can be due to a Ti3+ state ($Ti^{3+}$ → $Ti^{4+}$ + e$^-$). Such a presence of $Ti^{3+}$ also suggests the presence of oxygen vacancies or paramagnetic impurities.

Any variation from this value results from (i) a local crystal field from nearby atoms, (ii) spin-orbit coupling, and (iii) covalence or hybridization with neighbouring atoms. A g value of approximately 2.002 is very close to the free-electron value. This often indicates (i) the presence of an unpaired electron likely in an s-like orbital or a very lightly affected p-like orbitals, or (ii) minimal impact from heavy atoms, which would shift g more due to strong spin-orbit coupling. Otherwise, (iii) the electron is mostly located on a light element, like oxygen.[37–39] In oxide materials like $TiO_2$, a missing oxygen leaves behind two electrons that were shared with neighbouring Ti atoms during bonding. These electrons can be trapped at nearby cationic sites, forming $Ti^{3+}$ centres, or can self-trap at the vacancy site. These self-trapped vacancy sites are referred to as single-electron-trapped oxygen vacancies (SETOV).[40,41] For bulk SETOV, g ≈ 2.0025 to 2.0032 and is nearly isotropic, meaning g∥ ≈ g⊥. The features are typically sharp and narrow in nature, with a peak to peak line-width ($\Delta H_{pp}$) often less than 5 Gauss at X-band. They originate from localized electrons in F-centre-like bulk oxygen vacancies with minimal crystal field distortions. Such observations are seen in samples intentionally reduced by annealing in vacuum or $H_2$, or those that have undergone UV irradiation.[41]

On the other hand, an anisotropic distribution of g values is found for surface oxygen vacancies, with g∥ ≈ 2.0040 to 2.0055 and g⊥ ≈ 1.985 to 1.990.[42,43] These features are usually broader than bulk SETOV due to surface disorder and come from electrons trapped at an oxygen vacancy near the surface, interacting with surface Ti atoms and possibly other elements. Such features are present in UV-irradiated or chemically reduced $TiO_2$ that have been exposed to air or adsorbates. The unpaired electron in a SETOV is loosely bound in a nearly isotropic potential, similar to an F-centre in alkali halides.[44] It experiences little spin-orbit coupling and is not heavily hybridized with d-orbitals of Ti. Because of this, g remains close to the free-electron value of g = 2.0023. The SETOV is typically referred to as a "bulk defect" based on its location and environment. These defects are found deep within the crystal lattice and are distinct from surface oxygen vacancies, which are responsible for adsorption processes and have different g values. Bulk vacancies tend to be more symmetrical and stable, resulting in narrow lines, while surface vacancies are often more varied, resulting in broader or anisotropic EPR features.[45] In these samples both for pure and V-doped $TiO_2$ the dominant signal is centred at g ≈ 2.002. This strongly confirms that although these are nanoparticles, surface defects are less common in both samples. The most important conclusion from the above discussion is that the signature of oxygen vacancies or $Ti^{3+}$ ion formation is only at the bulk and is absent on the surface.

The hyperfine splitting comes from the magnetic interaction between the spin of an unpaired electron and the magnetic moment of a nucleus with non-zero nuclear spin (I). The hyperfine Hamiltonian is given by the Hamiltonian: $H_{HF} = S.A.I$, where S represents the electron spin, A is the hyperfine coupling constant and I stand for nuclear spin. A can have isotropic ($A_{iso}$) and anisotropic ($A_{dip}$) components. While $A_{iso}$ arises from Fermi contact interaction (electron density at the nucleus), $A_{dip}$ is due to dipole–dipole interaction between electron and nucleus. These interactions produce (2I + 1) equally spaced components that are also dependent on the hyperfine coupling constant (A). For the titanium $Ti^{3+}$ centers ($3d^1$) the coupling constants are $A_{iso}$ = −3.0 to −3.5 MHz and $A_{dip}$ = ~−2.0 to −2.5 for $^{47}Ti$ isotope, whereas for $^{49}Ti$ isotope the constants are $A_{iso}$ = −2.8 to −3.3 MHz and $A_{dip}$ = −1.9 to −2.3 MHz. On the contrary, for the Vanadium $V^{4+}$ centre the constants are $A_{iso}$ = 180–184 MHz, and $A_{dip}$ = 6–7 MHz. Hence the $A_{iso}$ of vanadium is much stronger than all the other components from which hyperfine splitting can be obtained.[46–49]

Titanium is naturally present in five isotopes: $^{46}Ti$, I=0, with abundance ~8.25 %, $^{47}Ti$, I = 5/2, with 7.44% abundance, $^{48}Ti$, I=0, with 73.72 % abundance, $^{49}Ti$, I = 7/2, with 5.41% abundance, and $^{50}Ti$, I = 0, with 5.18 % abundance. Hence only $^{47}$Ti and $^{49}$Ti have non-zero I, which can contribute to the hyperfine splitting. Therefore, only about ~ 13% of total titanium content contributes to the hyperfine splitting.[50] On the other hand, natural occurrence of vanadium is observed as $^{50}$V and $^{51}$V isotopes. Note that the $^{50}$V isotope is very rare ~0.25%, with I=6 whereas $^{51}$V isotopes are dominant in nature with ~99.75%, with I=7/2.

As discussed above the hyperfine splitting component in pure TiO$_2$ sample are only due to the Ti$^{3+}$ centres (3d$^1$) which are generally weak due to smaller values of coupling constant and thereby shows incomplete hyperfine structures.[51,52] With vanadium-doping the V$^{4+}$ centres (3d$^1$) present a strong hyperfine splitting due to the large A$_{iso}$ value. Hence, in the TV4 sample the most contribution to the hyperfine splitting comes from the $^{51}$V isotopes.[52]

Hence from the above analysis the following discussion is relevant to explain the connection between the finer details of electronic distribution of the Ti and V ions with the probable electro-chemical changes of these materials. EPR spectrum of TV4, exhibits a complex and highly anisotropic signal characteristic of V$^{4+}$ ions (d$^1$ configuration) in a solid matrix. The spectrum is dominated by a multi-line pattern, which is a direct consequence of the hyperfine interaction between the unpaired electron of V$^{4+}$ and the nuclear spin of the naturally abundant $^{51}$V isotope (I = 7/2). This interaction splits each electronic transition into 2I + 1 = 8 hyperfine lines.[52] The presence of multiple sets of hyperfine lines, even within the parallel and perpendicular regions, suggests that there might be more than one distinct V$^{4+}$ site within the TiO$_2$ matrix, or a distribution of local distortions around the V$^{4+}$ ions. This could be due to the varying proximity of oxygen vacancies, other dopants, or grain boundaries in the polycrystalline sample. While a dominant signal remains at g ≈ 2.002, additional resonances appear across the g = 1.94–2.20 range, indicative of multiple paramagnetic species. These include: (i) Ti$^{3+}$ centres (g ≈ 1.97–1.99), formed via partial reduction of Ti$^{4+}$ (ii) V$^{4+}$ species (g ≈ 1.96–2.05), reflecting the ESR-active d$^1$ configuration of vanadium ions (iii) oxygen-centred radicals such as O$^-$ or O$_2$•$^-$, stabilized on the surface via electronic interaction with vanadium dopants.[53] The substantial increase in ESR signal intensity for TV4 signifies a higher density of unpaired electrons. Such signatures may be correlated with enhancement of catalytic performance. This will be later discussed in the experimental section detailing the dark catalysis where ~51% RhB degradation was observed in TV4 sample. The presence of vanadium appears to induce oxygen vacancies and promote the formation of redox-active V$^{3+}$/V$^{4+}$/V$^{5+}$ centres, which facilitate electron trapping, interfacial charge separation, and reactive oxygen species (ROS) generation-all critical for efficient catalytic performance.[54]

Following the DFT, structural, UV-Vis and EPR studies to understand the structure correlated electronic properties, one need to understand the nature and role of the electronic nature of the surface of the materials. An XPS study provides evidence of the valence states and the degree of metal-oxygen covalence, all of which are essential for understanding the origin of the observed electronic behaviour.

The XPS survey spectra reveal the presence of C, Ti, V and O in all the doped samples and only C, Ti and O in the pure sample [Figure 6(a)]. The presence of C is due to adventitious Carbon, which comes from handling or even oil pollution inside the XPS chamber. All XPS binding energies were corrected by referencing the C-1s peak of adventitious carbon to 284.8 eV to compensate for surface charging effects [Figure SM4]. To analyse the presence of different oxidation states and different species of same element peak fitting analysis of high-resolution spectra of all elements have been carried out using the XPSPEAK41 software with the Shirley background and Pseudo-Voigt (weighted sum of a Gaussian and a Lorentzian function) type peak profile.

**Ti 2p spectrum:** The high-resolution Ti-2p XPS spectra for the pristine as well as V-doped TiO$_2$ samples consists two major peaks, which are corresponding to the spin-orbit split doublet ($2p_{3/2}$ and $2p_{1/2}$) of Ti$^{4+}$ with a splitting energy ($E_{SOS}$) of ~5.7 eV [Figure 6(b) and Supplementary material: Figure SM6]. For all the samples, these two peaks appear at ~ 458.5-459 eV for $2p_{3/2}$ and ~ 464.2-464.7 eV for $2p_{1/2}$ respectively, which is in good agreement with literature.[55,56] Note that the peaks are at slightly higher BEs ~458.8 eV for TV0 and TV1, but reduced by ~0.4 eV for higher doped samples. As the instrumental resolution is ~0.3 eV, this seems to be real shift of BEs for higher doped samples. The decrease of BEs of both peaks by ~0.4 eV in all highly doped samples, can be attributed to the enhanced electron density around Ti introduced by the n-type V dopant, which is also confirmed from the spin-density plot for TV0 system [Figure 5(c)], obtained from ab-initio calculation. This increased electron sharing between Ti and O strengthens the Ti–O covalence, leading to a shorter and stronger Ti–O bond, as further supported by the XRD analysis and Raman spectroscopy study (blue-shift of Raman peak position due to V doping).

The presence of Ti$^{3+}$ in TV0 has been confirmed from the EPR studies. However, EPR studies provided information on the presence of oxygen vacancies and Ti$^{3+}$ formation probabilities in the bulk but not on the surface. Hence, XPS being a surface analysis does not reveal the presence of Ti$^{3+}$ and reveals only the signature of Ti$^{4+}$ [Figure 6(b)].

**V 2p spectrum:** The high-resolution V 2p spectra for the V-doped TiO$_2$ samples were de-convoluted into six peaks corresponding V$^{+3}$, V$^{+4}$, and V$^{+5}$ oxidation states [Figure 6(f) and supplemental material: SM7]. Each oxidation states gives two peaks ($2p_{3/2}$ and $2p_{1/2}$) due spin-orbit splitting, with an energy separation ($E_{SOS}$) of ~7.5 eV. All peak positions and fitting parameters are listed (supplemental material: Table SM6). The $2p_{3/2}$ peaks appear at ~515-515.6 eV for V$^{3+}$, ~516-515.6 eV for V$^{4+}$, and ~517-515.5 eV for V$^{5+}$, while the $2p_{1/2}$ peaks lie at ~522.5-523.1 eV, ~523.5-524.1 eV, and ~524.5-525.1 eV, respectively.[57,58] The binding energies for sample TV1 are slightly higher than those of the other doped samples. The relative concentrations of each oxidation states were quantified from the integrated areas of the respective peaks, defining the fraction of $V^{5+}$ as: $f_{V^{5+}} = \frac{area(V^{5+})}{area(V^{3+})+area(V^{4+})+area(V^{5+})}$. From the results it was observed that with increasing concentration of V-doping, $f_{V^{5+}}$ decreases from 0.18 in TV1, to 0.08 in TV2, then increases again to 0.16 in TV3 and 0.17 in TV4. The presence of V in multiple oxidation states makes it a highly active redox species: $V^{3+}$ and $V^{4+}$ can be oxidized to $V^{4+}$ and $V^{5+}$, while $V^{4+}$ and $V^{5+}$ can be reduced back to $V^{3+}$ and $V^{4+}$. This multi-valence enables efficient activation of $H_2O_2$ to generate hydroxyl radicals (·OH). As a result, V-doped $TiO_2$ (TVO) acts as an effective catalyst for both dark-Fenton and photo-Fenton like reactions.

**O 1s spectrum:** The O 1s XPS spectra for all samples [Figure 6(c and e) and supplementary material: Figure SM5] reveals one single feature composed of two convoluted peaks, one attributed to lattice oxygen ($O_L$) in the range 529.7 eV to 530.2 eV, while the other in the range 530.8 eV to 531.1 eV can be reasonably attributed to V-O and carbonate species (i.e. Ti/V-O-C and C=O bonds) or to surface oxygen species.[59] Because the XPS measurements were performed on the Ar-etched surface, the contribution from any adsorbed oxygen species can be excluded. The V–O bond is more covalent than the Ti–O bond because V is more electronegative than Ti. This increased covalence reduces the electron density around oxygen in the V–O environment, resulting in a slight increase in the binding energy of the O 1s core-level electrons. To examine oxygen non-stoichiometry in the samples, the oxygen peak intensity was normalized to the maximum Ti peak intensity in the high-resolution XPS spectra.[60] The normalized oxygen intensity for all the doped samples is more in comparison with the pristine system, indicating higher lattice oxygen content and fewer oxygen vacancies (a typical point defects in transition-metal oxides) as a result of V incorporation [Figure 6(g)]. The normalized oxygen intensity decreases slightly from TV1 to TV2, then increases steadily for TV3 and TV4. This trend is consistent with the increasing amount of $V^{5+}$ [Figure 6(h)], as the lattice pulls in additional oxygen to compensate for the extra positive charge, thereby reducing oxygen vacancies. Although fewer oxygen vacancies normally lead to an expansion of the unit cell, the strong V-O covalence introduced by doping causes an overall lattice contraction, which is reflected in the observed decrease in unit-cell volume with increasing doping concentration [Figure 1(c)].

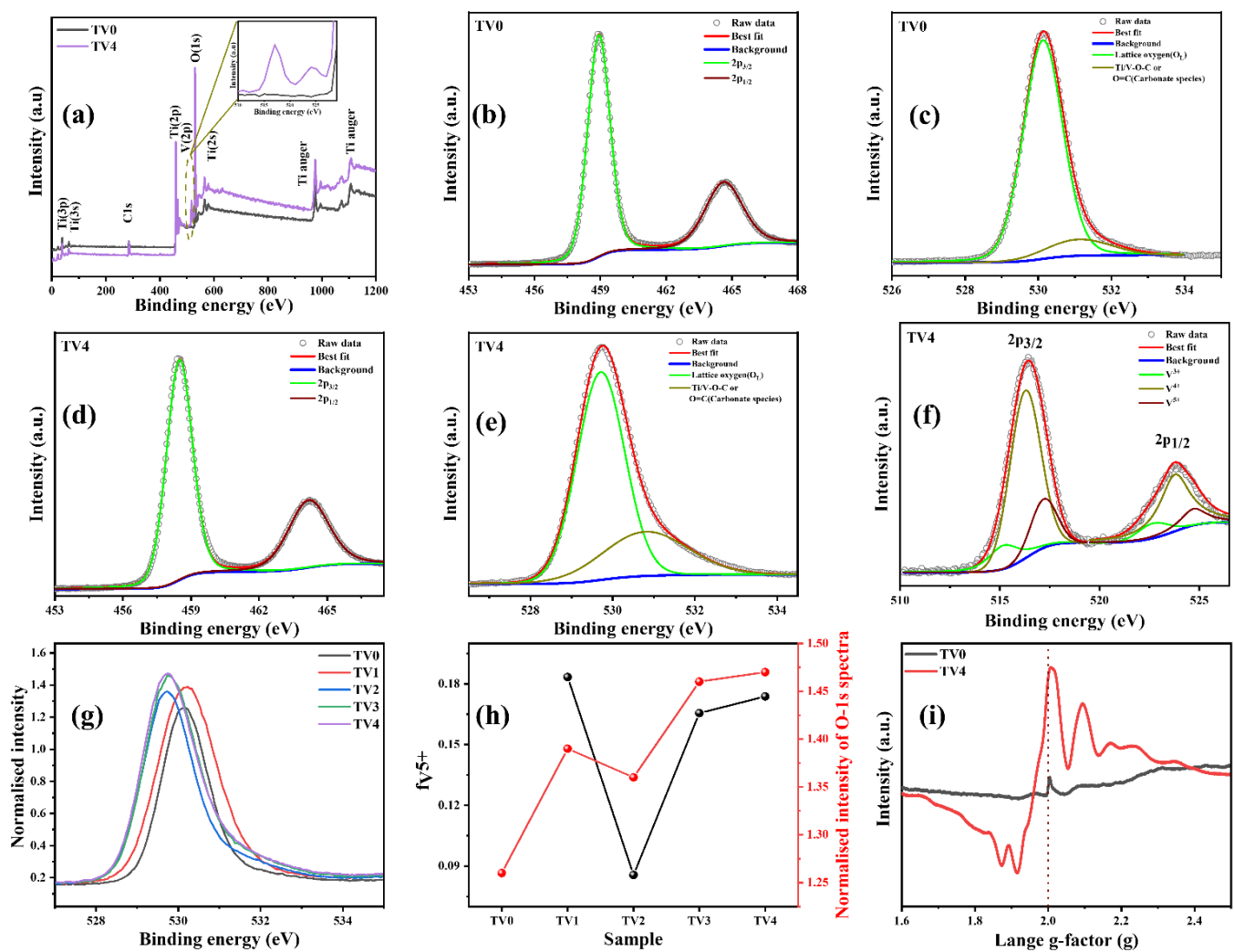

Figure 6: (a) XPS survey spectra of TV0 and TV4 samples. High-resolution spectra of (b) Ti 2p, and (c) O 1s for sample TV0 and (d) Ti 2p, (e) V 2p, and (f) O 1s for sample TV4. (g) Variation in the normalized oxygen spectral intensity and (h) the fractional area of $V^{5+}$ ($f_{V^{5+}}$) and normalised intensity of O-1s spectra as a function of doping concentration. (i) ESR spectra of TV0 and TV4.

### 3.4. Catalytic activity:

Often it is observed that catalytic activity of a particular material is evaluated by addition of $H_2O_2$ which acts as a reducing/oxidizing agent.[61] The breakdown of $H_2O_2$ is an important factor in the catalytic activity; $H_2O_2 \rightarrow$ •OH + OH⁻, where •OH is the hydroxyl radical which is usually formed as a transient, high-energy species during oxidation processes. The •OH radical and the OH⁻ ion are chemically and physically very different because of their electronic structure and reactivity. The •OH radical is neutral and contains one unpaired electron and therefore is paramagnetic in nature whereas, the OH⁻ ion is negatively charged and hence the electrons are paired thereby having diamagnetic nature.[62] While •OH radicals are extremely unstable and highly reactive, the OH$^-$ ion is stable in aqueous solution under basic conditions. Fe$^{2+}$ catalyses the decomposition of H$_2$O$_2$ to generate highly reactive •OH radicals through the Fenton reaction, while other transition-metal ions such as Cu$^+$ and Mn$^{2+}$ can induce similar oxidative processes, commonly referred to as Fenton-like reactions.[63] TiO$_2$ is well known to facilitate catalysis either with the help of a photon or by presence of Ti$^{3+}$.

The generation of •OH is one of the most significant descriptors to evaluate the performance of H$_2$O$_2$-based Fenton and Fenton-like catalysis. We thereby qualitatively investigated the •OH generation by TiO$_2$ and V-doped TiO$_2$ (V: TiO$_2$) catalysts following scavenger test while degrading RhB dye. The interaction between Ti/V sites and H$_2$O$_2$ molecules is pivotal for understanding the origin of the Fenton-like catalytic performance and how it is improved by vanadium doping. By revealing the work function,[64] adsorption energy ($E_b$),[65] and charge transfer analysis,[66] the increase of H$_2$O$_2$ activation upon V substitution is clearly evident from the theoretical analysis. In this respect, a systematic calculation of these two depicters at the main facet of anatase-TiO$_2$, i.e., (101) in TiO$_2$ and V: TiO$_2$ was evaluated. The (101) growth direction was extracted from the XRD patterns with high peak intensity [Figure 1(a)].

The electronic structure of the TiO$_2$ slab, exposing the (101) facet, shows a band gap of 2.54 eV [Figure 7(a)], which reduces to 2.46 eV upon V doping with a defect level at ~ 0.59 eV above the valence band, mainly dominated by V-3d states [Figure 7(b)]. The planar average potential analysis of the examined facet shows a work function value of 7.23 eV, which lowers to 6.52 eV upon V-incorporation [Figure 7(c and d)]. This is the macroscopic manifestation of charge transfer from the catalyst to the reactant, facilitated by V-doping. In the following sections, further investigation has been carried out to obtain the atomistic essence.

The generation of •OH includes H$_2$O$_2$ adsorption and consequent reactions as follows: * (i) → *H$_2$O$_2$ (ii) →*OH+•OH (iii) →*+2•OH (iv) [Figure 7(i)], where * indicates the adsorption state.[67] Figure 7(e) and Figure 7(g) illustrates the accumulation of the electron region between the OH group of H$_2$O$_2$ molecules and Ti/V site. This implies that the oxygen donates electrons to the Ti/V atom to form a dative bond between oxygen and the catalytically active site. This is confirmed from the enhancement of Bader charge of both Ti and V-atom after H$_2$O$_2$ adsorption. As illustrated in [Figure 7(i)], the corresponding $E_b$ of H$_2$O$_2$ on the V-sites was determined to be -0.77 eV, which was lower than that of the Ti site (-0.72 eV), proving the supremacy of V over Ti towards H$_2$O$_2$ adsorption.

To further understand the redox reaction before and after $H_2O_2$ decomposition on the (101) surface, we calculated the Bader charge and magnetic moments of Ti/V ions and the fragments of $H_2O_2$ during the reaction [Figure 7(j)]. The calculated magnetic moment values of Ti/V ions are similar before and after the adsorption of $H_2O_2$, while Bader charge values increase slightly. After the adsorption of $H_2O_2$, the calculated Bader charge values of Ti and V are around 1.98|e| and 3.07|e|, respectively, and the calculated magnetic moment values of both Ti and V atoms are around 0 $\mu_B$ and 1.06 $\mu_B$, respectively. The results indicate that the oxidation state of Ti and V atoms upon $H_2O_2$ adsorption is $Ti^{4+}$ and $V^{4+}$, respectively. Furthermore, after the $H_2O_2$ decomposition to •OH, the calculated Bader charge value of both Ti and V atoms become 1.96|e| and 2.96|e|, respectively, and the calculated magnetic moment of both Ti and V atoms become 0.042 $\mu_B$ and 0 $\mu_B$, respectively, as listed in [Figure 7(j)]. The changes in Bader charge and the magnetic moment suggest that the V is oxidized from $V^{4+}$ to $V^{5+}$, while the Ti is maintained at $Ti^{4+}$. The calculated results indicate that $H_2O_2$ on the V site will be decomposed to OH radical, while it is easier for Ti site to reduce it to $OH^-$ ion. As a result, we predict that the redox reaction of $V^{4+}$ follows the initiation step of Haber–Weiss mechanism for Fenton-like activity.[68] A similar kind of mechanism can also be followed by $V^{3+}$ and $Ti^{+3}$ as discussed in previous studies.[4,69,70] Thus, V-substituted $TiO_2$ shows better catalytic activity by degrading RhB dye solution, which is mainly dominated by the Fenton-like reaction of $V^{+4}$.

The changes in the free energies for •OH formation in the (101) facet of $TiO_2$ and V: $TiO_2$ are displayed in [Figure 7(i)]. The $H_2O_2$ adsorption (i) step is exothermic in nature for both cases, while the generation of •OH radical (ii) was observed as the rate-limiting step due to the highest energy barrier. Notably, $H_2O_2$ adsorption to V was thermodynamically more favourable than to Ti ($E_b$ = -0.77 eV for V-sites and -0.72 eV for Ti sites). Subsequently, the $H_2O_2$ on the catalyst surface splits into *OH, accompanied by one •OH desorption. The energy barriers required for $H_2O_2$ split and •OH desorption on the (101) facet of $TiO_2$ was found to be 3.11 eV, which is much higher than for the V site (1.96 eV), demonstrating the superior activity of V: $TiO_2$ for $H_2O_2$ activation. It is therefore expected that the $V^{+4}$ sites outperform $Ti^{+4}$ sites in terms of $H_2O_2$ activation. It is expected that $Ti^{+3}$ sites can show better activity than $Ti^{+4}$ towards $H_2O_2$ activation, but its effect can't outperform $V^{+4}$ due to the relatively smaller concentration of $Ti^{+3}$ as confirmed from the EPR study. As discussed earlier, the generation of $2^{nd}$ •OH radicals could not be possible for the V site due to a very high energy barrier, while it is energetically favourable for Ti. So, the OH group on the V site will be reduced to the OH⁻. Thus, the DFT study established the proposed Fenton-like reaction mechanism for $V^{+4}$ and the same analogy can be drawn for $Ti^{3+}$ and $V^{3+}$ as well, and confirms why V-doping in anatase-TiO2 is beneficial for non-illuminated catalysis as compared to pure TiO2.

$$Ti^{3+} + H_2O_2 \rightarrow Ti^{4+} + HO\cdot + OH^-$$

$$Ti^{4+} + H_2O_2 \rightarrow Ti^{3+} + HOO\cdot + H^+$$

$$V^{3+} + H_2O_2 \rightarrow V^{4+} + HO\cdot + H^+$$

$$V^{4+} + H_2O_2 \rightarrow V^{3+} + HOO\cdot + H^+$$

$$V^{4+} + H_2O_2 \rightarrow V^{5+} + HO\cdot + OH^-$$

$$V^{5+} + H_2O_2 \rightarrow V^{4+} + HOO\cdot + H^+$$

$$2H_2O_2 \rightarrow HO\cdot + HOO\cdot + H_2O$$

Having established the presence of multiple valence state of V in TiO₂ lattice from XPS studies and finding a theoretical explanation of how this may enhance the catalytic properties, it is now important to verify this experimentally.

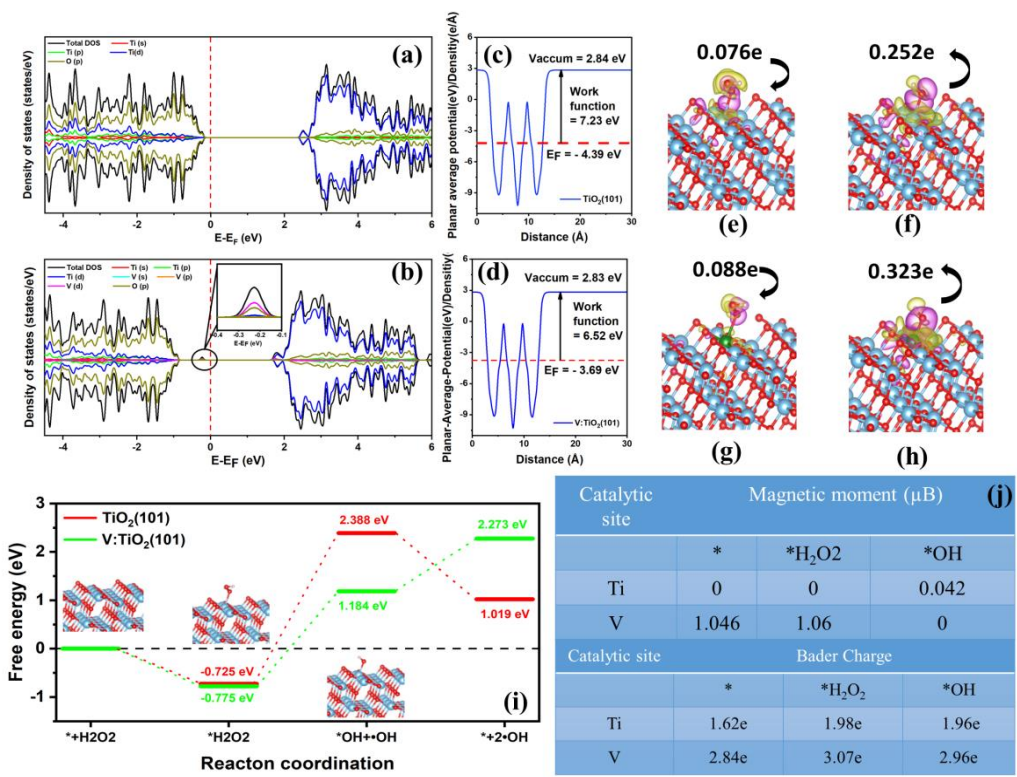

Figure 7: Orbital resolved density of states plots for slabs exposing (101) surface of (a) $TiO_2$ (b) V-doped $TiO_2$ (Fermi energy is set at 0 eV). Planer average potential along [101] representing the work function of (101) facet of (c) $TiO_2$ (d) V-doped $TiO_2$. Calculated Bader charge transferred from slab in the intermediate steps and charge density difference plots (Isosurface level 0.001 eV/Å$^{-3}$) between catalyst and reactant for *$H_2O_2$ ((e) and (f)), *OH ((g) and (h)) (Solid spheres having colour of blue, green, red and white represents Ti, V, O and H atom respectively. The yellow and pink surfaces represent charge accumulation and depletion). (i) Calculated free energy profiles for H2O2 activation and •OH formation on (101) facet of $TiO_2$ and V-doped $TiO_2$. (j) Calculated magnetisation and Bader charge of catalytic active site i.e. Cu ($TiO_2$) and V (V-doped $TiO_2$) atoms for each intermediate steps.

### 3.4.1. Performance of doped and un-doped $TiO_2$ towards $H_2O_2$ activation for wastewater treatment:

The Fenton-like performance of the $Ti^{3+}$, $V^{3+}$, and $V^{4+}$ as described above, can activate $H_2O_2$ into reactive •OH radicals and $OH^-$ ions under dark conditions and therefore can substantially degrade RhB dye. A comparative study of the Fenton-like performance of the $TiO_2$ and V-$TiO_2$ systems were carried out. The degradation efficiency, η (%), was calculated using the equation: $\eta(\%) = \left(\frac{A_t - A_0}{A_0}\right) \times 100$, where $A_0$ is the absorbance of the dye solution before degradation and is proportional to concentration (C) of the dye solution. $A_t$ is the absorbance of the dye solution at a time t [supplemental material: Figure SM8]. The system containing only $H_2O_2$ without any catalysts, degraded 3.8 % dye in 150 min. This is almost negligible and can be due to the thermal activation of $H_2O_2$.[71] The η (%) observed for the pure TV0/$H_2O_2$ system is higher ~7.6 % than $H_2O_2$. As V-doping concentration increased, η (%) increased to ~14.45% for TV1, ~23.9% for TV2, ~25.9% for TV3, and 51.2% for TV4 [Figure 8(c)]. Hence the TV4/$H_2O_2$ system exhibits maximum performance. From XPS studies it was found that vanadium appears as redox-active $V^{3+}/V^{4+}/V^{5+}$ centres. Hence, the enhancement in performance can be directly correlated with the availability of $V^{4+}/V^{5+}$ pairs, affirming the central role of vanadium doping in enabling the Fenton-like catalytic mechanism. Figure 8(a) represents the time evolution of the relative dye-concentration $(\frac{C_0}{C_t})$ which is proportional to $\frac{A_0}{A_t}$, where $C_0$ is the concentration of dye solution at time t=0 and $C_t$ are the same at time t. The logarithmic change in relative concentration was found to follow exponential linear relation with time

[Figure 8(b)]. Hence, the relation can be expressed as: $\ln\left(\frac{C_0}{C_t}\right) = kt$, where, $k$ represents the rate constant. Therefore, $k = \frac{1}{t} \ln \frac{C_0}{C_t}$. This signifies a first-order reaction kinetics. The degradation percentages and calculated rate constant for all the catalysts has been listed in Supplemental Material: Table SM7.

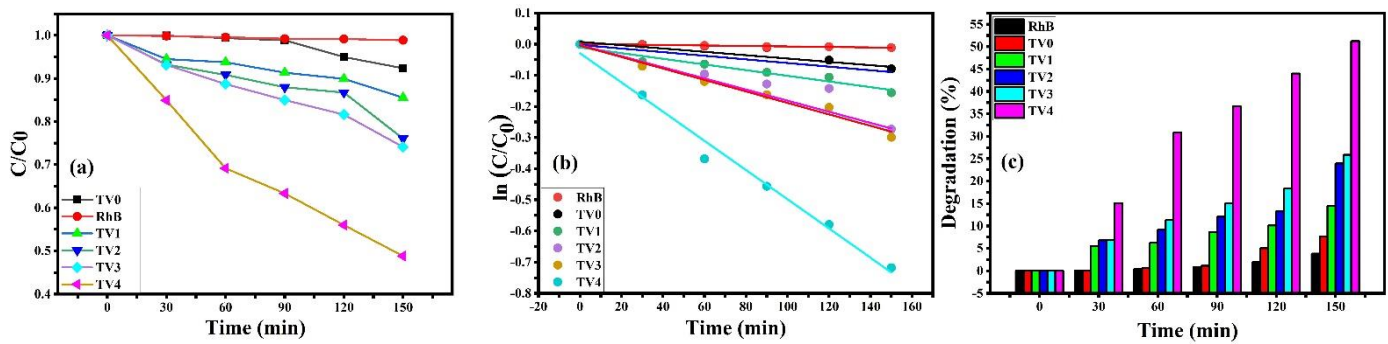

Figure 8: (a) Change in the relative concentration of the dye solution over time. (b) The logarithmic plot of the relative concentration of the dye solution over time and the linear nature confirms the 1$^{st}$ order reaction kinetics. (c) The dye degradation percentage over time.

### 3.4.2. Role of Hydroxyl Radical: Scavenger Test using tert-butanol (t-BuOH, (CH3)3COH):

To evaluate the contribution of •OH in the catalytic degradation of RhB, scavenger experiments were conducted using t-BuOH as scavenger, a selective •OH radical quencher.[72] The experiment was performed using 30 mL of RhB solution for both the TV0 and TV4 samples. For each test, 30 mg of the sample was added separately to the RhB solution, followed by the addition of 6 mL of $H_2O_2$ (30% V/V) and 2 mL of t-BuOH. The degradation efficiency η (%) was monitored for 150 minutes under dark conditions.

Upon the addition of t-BuOH, the η (%) for TV0/$H_2O_2$ was suppressed to ~1.5% from ~7.8%, and for TV4/$H_2O_2$ was strongly suppressed to ~1.5% from ~51.2% in 150min [supplemental material: Figure SM9]. t-BuOH is well known to quench •OH without affecting other ROS pathways.[73] Hence, this strong inhibition of RhB degradation upon t-BuOH addition, confirms that •OH is the primary and essential oxidative species involved in the catalytic reaction for both catalysts. This also provides strong evidence that alternative species such as superoxide (•$O_2^-$) play negligible role in the degradation of RhB for the TV0 and TV4 catalysts. These findings are consistent with earlier reports on TiO$_2$-based photo-catalysis where hydroxyl radicals were identified as the key oxidizing agents.[74]

### 3.4.3. Re-usability test:

To check the stability and reusability of the nanoparticles for degrading RhB dye, three cycles of the degradation process were conducted. After every cycle, the samples were extracted using centrifugation, washed with distilled water and ethanol, and then dried in a microwave oven. After drying, these samples were utilized again for the catalytic dye degradation experiment, keeping every experimental parameter (i.e; temperature, concentration of dye and the time for degradation) similar. To confirm structural stability, XRD analysis was performed, which showed the retention of the anatase phase, indicating that the nanoparticles maintained their structural integrity after the degradation process [Supplemental Material: Figure SM10(a)]. However, the broadening of the XRD peaks can be attributed to a decrease in crystallinity [supplemental material: Figure SM10(b)]. The degradation percentage decreases by ~0.55 % per cycle for TV0 and ~1.1% per cycle for TV4 [Supplemental Material: Figure SM10(c and d)]. However, this decrement in performance may be due to the lesser amount of retrievable catalyst than the originally used. This was due to the loss of nanomaterial during the filtration and washing processes. Hence, it is most likely that the efficiency of the catalyst does not vary significantly upon reusage.

## 4. Summary:

Pure and V-doped $TiO_2$ catalysts has been synthesised successfully following sol-gel wet chemical synthesis route. XRD and Raman spectroscopy study confirms that the crystallinity of the catalysts to be predominantly of the anatase (space group: $I4_1/amd$) phase. Despite the negligible decrement of oxygen vacancy, lattice contraction was observed with increasing doping concentration, mainly attributed to the enhanced metal-oxygen covalence and therefore reduction in bond length. The enhanced covalence facilitate better charge transfer property. Lattice contraction generates lattice strain, which increases with doping resulting in reduction of crystallite size. The BET analysis confirms the drastic enhancement in surface porosity and therefore in the specific surface area in the doped samples, which make the V-doped $TiO_2$ a better catalyst with large reactive surface area. The presence of V in the $TiO_2$ lattice and oxygen vacancies create redox-active $V^{3+}/V^{4+}$ and $V^{4+}/V^{5+}$ centres as found from XPS and EPR studies. A signature of the presence of $Ti^{3+}$ as confirmed from EPR suggests the formation of redox active $Ti^{3+}/Ti^{4+}$ centre as well, although the concentration of $Ti^{3+}$ is very less. The electronic structure analysis from DFT study suggests reduction in band-gap energy and emergence of non-bonding V-3d electron as localised mid-gap state. The enhanced metal-oxygen covalence upon V-doping and the mid-gap state are the possible reason for the electronic structure renormalisation. UV-Vis spectroscopy provides the experimental evidence for the same. The increasing Urbach energy signifies the enhanced lattice disorder attributed to difference in crystal radii, modification in electronic interaction and point defects. The localised V-3d electron increases the density of states near the Fermi energy, which leads to better adsorption of $H_2O_2$. In particular the redox-active $V^{4+}/V^{5+}$ centres are known to facilitate electron trapping, interfacial charge separation, and reactive oxygen species (ROS) generation-all critical for efficient catalysis process. From catalytic experiments the efficiency, η (%) increased drastically from ~7.6% for un-doped TV0 to ~14.45% for TV1, ~23.9% for TV2, ~25.9% for TV3, and 51.2% for TV4, implying the strong contribution of the redox-active $V^{4+}/V^{5+}$ centres. Theoretical calculations reveal that the $V^{+4}$ sites outperform $Ti^{+3}$ sites in terms of $H_2O_2$ activation, confirming the experimental results. The role of OH radical was found to be the prominent factor in the dark-catalytic degradation of RhB dye as compared to other ROS like superoxide ($•O_2^-$) or photo-generated holes ($h^+$) etc. The reusability of the catalyst was found to be excellent with a very nominal reduction in efficiency.

5. Conclusions:

The enhanced catalytic performance of V-doped $TiO_2$ arises from the synergistic effects of structural modulation, electronic property tailoring, enhance reactive surface area, redox-active surface chemical states, and the dynamic activation of $H_2O_2$ via a Fenton-like redox pathway. The incorporation of $V^{4+}$ ions into the $TiO_2$ lattice has been shown to significantly enhance its catalytic activity and visible light absorption. These findings and in-depth analysis highlight the critical importance of dopant-induced surface chemistry over mere physical surface area in designing highly efficient dark catalysts.

**Associated Content**

Supplemental material.

**Corresponding Author**

*Email id: sens@iiti.ac.in

**Author Contributions**

**Manju Kumari:** Conceptualization, Methodology, Data curation, Investigation, Writing – original draft. **Dilip Sasmal:** Methodology, DFT calculation, Investigation and Interpretation, Writing – original draft, review & editing. **Suresh Chandra Baral:** Data curation, Formal analysis and Validation, Writing – review & editing. **P. Maneesha:** Formal analysis and Validation. **Poonam Singh**: Formal analysis. **Abdelkarim Mekki:** Data curation. **Khalil

**Harrabi:** Data curation. **Somaditya Sen:** Conceptualization, Fund acquisition, Project administration, Supervision, Formal analysis and Validation, Writing – review & editing.

‡These authors contributed equally.

**Declaration of competing interest:**

The authors declare that they have no known competing financial interests or personal relationships that could have appeared to influence the work reported in this paper.

**Acknowledgement:**

The authors MK and DS would like to thank the University Grants Commission, India for fellowship. SCB thanks DST INSPIRE, PM thanks PMRF, India and PS thanks IIT Indore for providing fellowship. The Authors acknowledge the Department of Science and Technology (DST), Govt. of India for providing the funds (DST/TDT/ AMT/2017/200). The authors thank the Sophisticated Instrument Centre (SIC), IIT Indore for access to the BET surface area analyser facility. The authors extend their appreciation to the Department of Science and Technology (DST), Govt. of India, for allocating a FIST instrumentation fund to the Department of Physics at IIT Indore to purchase a Raman Spectrometer (Grant Number SR/FST/PSI-225/2016). The authors acknowledge Ms. Manisha Yadav from Delhi University for EPR measurement. A.M, K.H, and S.S acknowledge the support of the King Fahd University of Petroleum and Minerals, Saudi Arabia, under Grant No. DF191055 DSR project.

# Increased Covalence and V-center mediated Dark Fenton-Like Reactions in V-doped TiO2: Mechanisms of Enhanced Charge-Transfer

*Manju Kumari[1‡], Dilip Sasmal[1‡], Suresh Chandra Baral[1], Maneesha P[1], Poonam Singh[1], Abdelkerim Mekki[2,3], Khalil Harrabi[2,3], Somaditya Sen[1]\**

[1]Department of Physics, Indian Institute of Technology Indore, Indore, 453552, India

[2]Department of Physics, King Fahd University of Petroleum and Minerals, Dhahran 31261, Saudi Arabia

[3]Center for Advanced Materials, King Fahd University of Petroleum and Minerals, Dhahran 31261, Saudi Arabia

*Corresponding authors: sens@iiti.ac.in

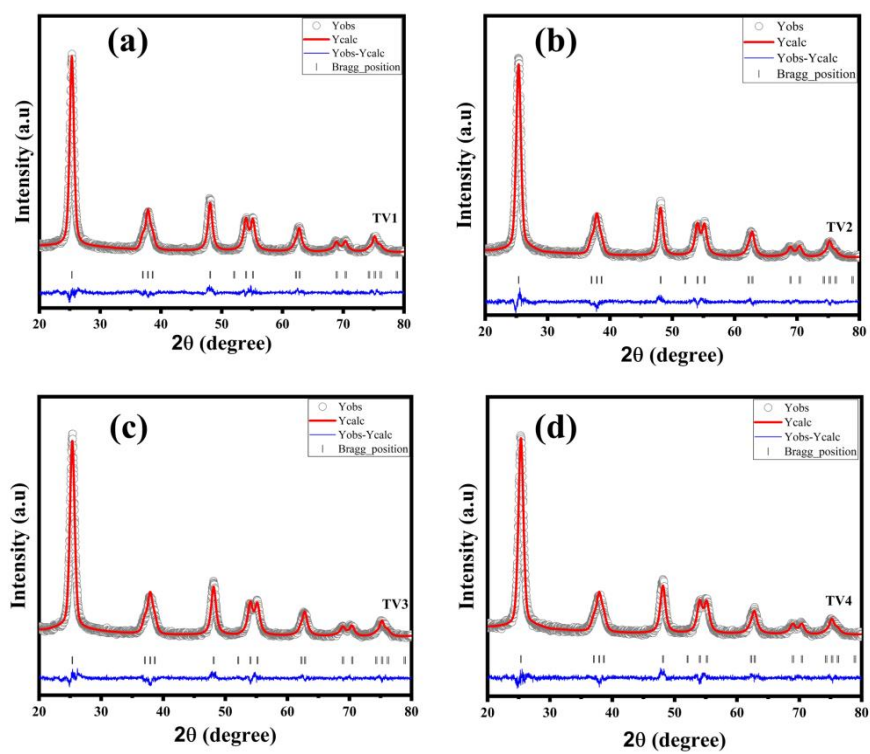

*Figure SM1: Reitveld refined pattern for samples: (a) TV1, (b) TV2, (c) TV3, and (d) TV4 (grey bars indicate Bragg peak positions).*

*Table SM1: Reitveld refinement parameters demonstrating the goodness of fit.*

| Samples | $R_p$ | $R_{wp}$ | $R_{exp}$ | $\chi^2$ |
|---|---|---|---|---|
| TV0 | 13.3 | 13.7 | 10.91 | 1.59 |
| TV1 | 10.9 | 11.1 | 9.05 | 1.51 |
| TV2 | 11.7 | 11.4 | 9.36 | 1.49 |
| TV3 | 11.9 | 11.7 | 9.38 | 1.57 |
| TV4 | 12.5 | 12.1 | 9.87 | 1.51 |

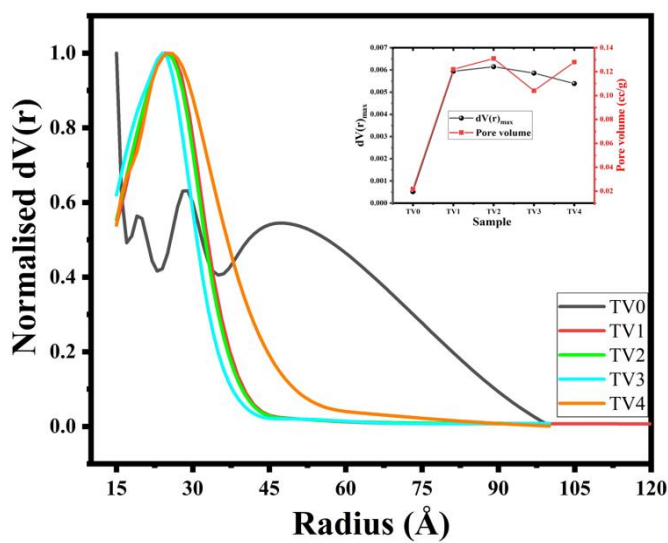

*Figure SM2: A Berrett-Joyner-Halenda (BJH) pore size distribution obtained from the $N_2$ adsorption-desorption isotherm. (Inset: variation of total pore volume with increasing doping concentration)*

*Table SM2: Surface area and pore structure parameters calculated from BET analysis.*

| Sample | BET Constant (C) | Monolayer Capacity (mmol/g) | Specific Surface Area (m²/g) | Pore Radius (Å) | Pore volume(cc/g) |
|---|---|---|---|---|---|
| TV0 | 10 | 2.26 | 10.64 | 15.31 | 0.022 |
| TV1 | 70 | 21.54 | 106.57 | 24.64 | 0.122 |
| TV2 | 98 | 21.08 | 169.9 | 24.58 | 0.131 |
| TV3 | 100 | 18.33 | 85.89 | 24.66 | 0.104 |
| TV4 | 103 | 19.34 | 97.78 | 24.56 | 0.128 |

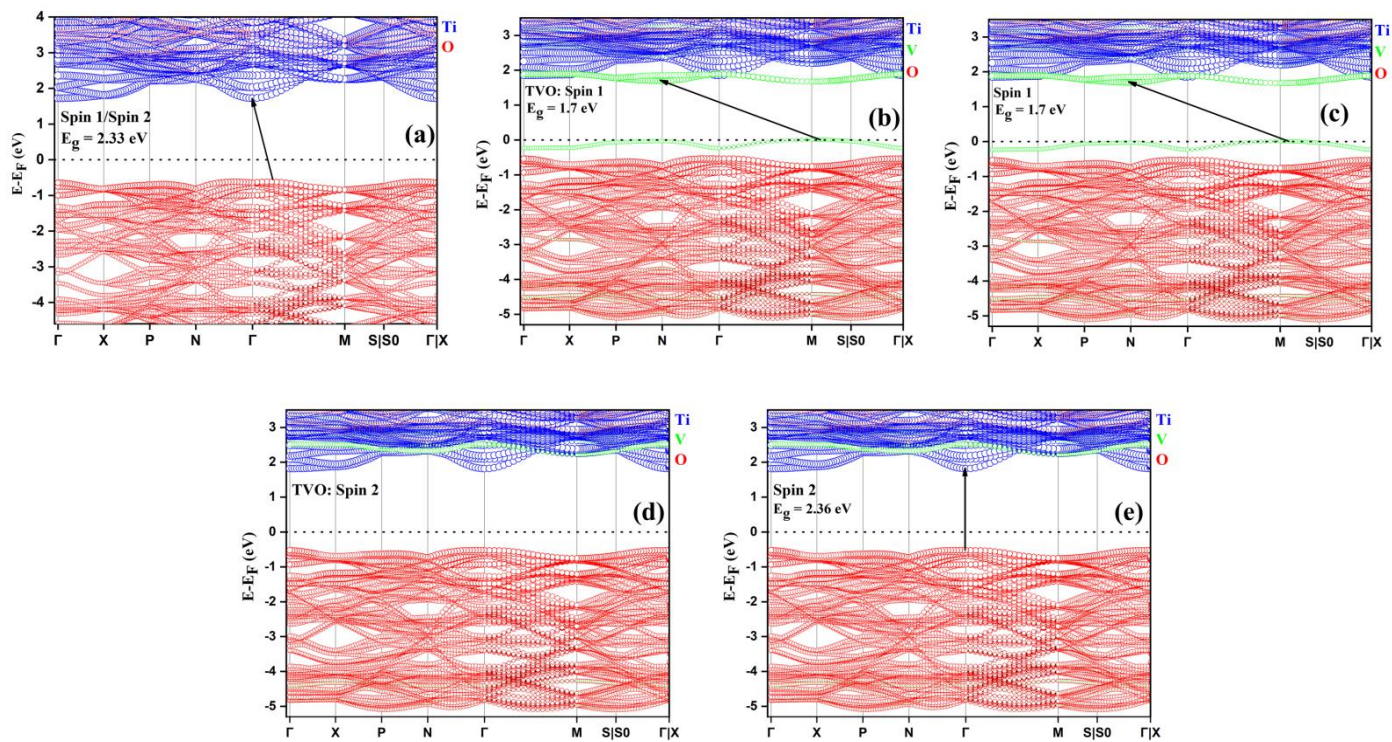

*Figure SM3: Calculated spin-polarised band structure and corresponding band gaps for the 2×2×1 supercell of: (a) pristine $TiO_2$, (b, d) V-doped $TiO_2$, and (c, e) V-doped $TiO_2$ with oxygen-vacancy ($O_V$).*

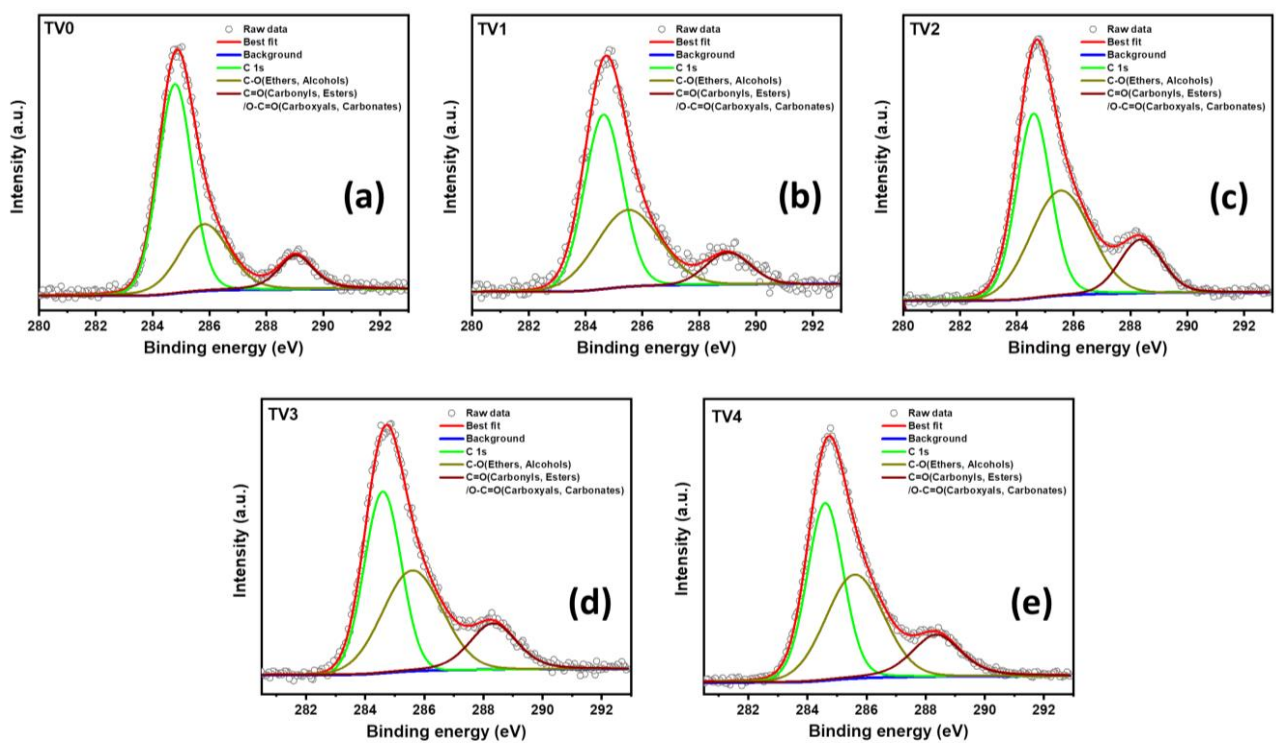

*Figure SM4: High resolution XPS spectra of C 1s for sample (a) TV0 (b) TV1 (c) TV2 (d) TV3 (e) TV4 with peak de-convolution illustrating the contribution from different chemical surroundings.*

| Samples | Peak 1 | | | | Peak 2 | | | | Peak 3 | | | | $\chi^2$ |
|---|---|---|---|---|---|---|---|---|---|---|---|---|---|
| | B.E. (eV) | FWHM (eV) | Area | G-L | B.E. (eV) | FWHM (eV) | Area | G-L | B.E. (eV) | FWHM (eV) | Area | G-L | |
| TV0 | 284.79 | 1.41 | 6669.524 | 9 | 285.82 | 2.03 | 3235.336 | 22 | 289.04 | 1.46 | 1315.816 | 46 | 1.509 |
| TV1 | 284.63 | 1.58 | 4219.139 | 7 | 285.49 | 2.64 | 3064.193 | 5 | 289.01 | 1.84 | 860.897 | 0 | 1.620 |
| TV2 | 284.59 | 1.42 | 11387.889 | 14 | 285.54 | 2.48 | 10622.77 | 0 | 288.36 | 1.75 | 4066.829 | 14 | 1.534 |
| TV3 | 284.60 | 1.47 | 5870.957 | 0 | 285.59 | 2.38 | 5372.491 | 4 | 288.33 | 1.87 | 2181.842 | 31 | 1.482 |
| TV4 | 284.59 | 1.44 | 13004.12 | 11 | 285.59 | 2.30 | 11854.92 | 6 | 288.36 | 1.95 | 4684.229 | 39 | 1.421 |

*Table SM3: Fitting parameters for XPS spectra of C-1s*

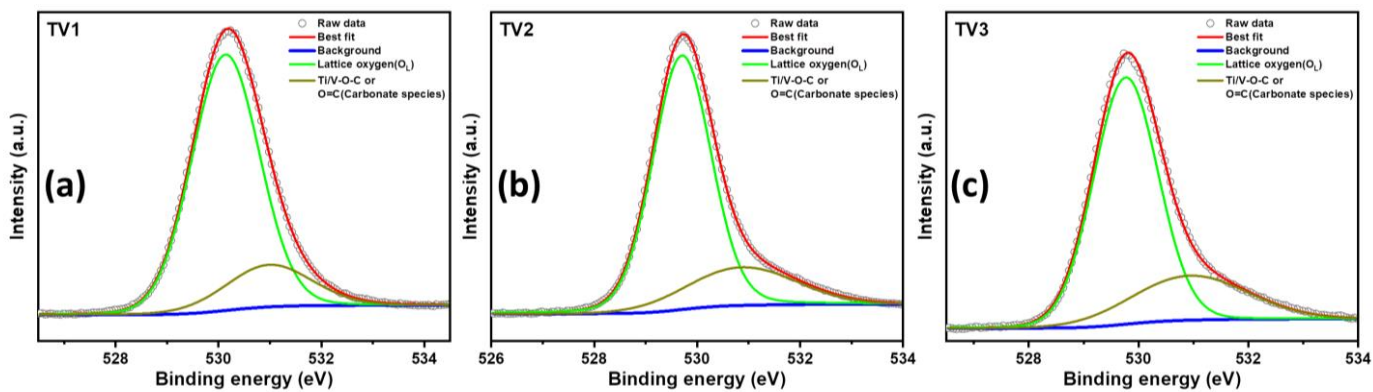

*Figure SM5: High resolution XPS spectra of O 1s for sample (a) TV1 (b) TV2 (c) TV3 with peak de-convolution illustrating the contribution from different chemical environment.*

*Table SM4: Fitting parameters for XPS spectra of O-1s*

| Samples | Peak 1 | | | | Peak 2 | | | | $\chi^2$ |
|---|---|---|---|---|---|---|---|---|---|
| | B.E. (eV) | FWHM (eV) | Area | G-L | B.E. (eV) | FWHM (eV) | Area | G-L | |
| TV0 | 530.13 | 1.27 | 139324.6 | 13 | 531.11 | 2.09 | 16122.71 | 0 | 4.392 |
| TV1 | 530.14 | 1.55 | 154033.5 | 6 | 530.9 | 1.96 | 33017.9 | 12 | 3.396 |
| TV2 | 529.71 | 1.34 | 219525.3 | 12 | 530.87 | 2.5 | 58238.66 | 0 | 5.177 |
| TV3 | 529.77 | 1.38 | 111837.7 | 5 | 530.93 | 2.48 | 35337.34 | 0 | 5.405 |
| TV4 | 529.71 | 1.37 | 199047.4 | 4 | 530.8 | 2.44 | 76181.47 | 5 | 12.708 |

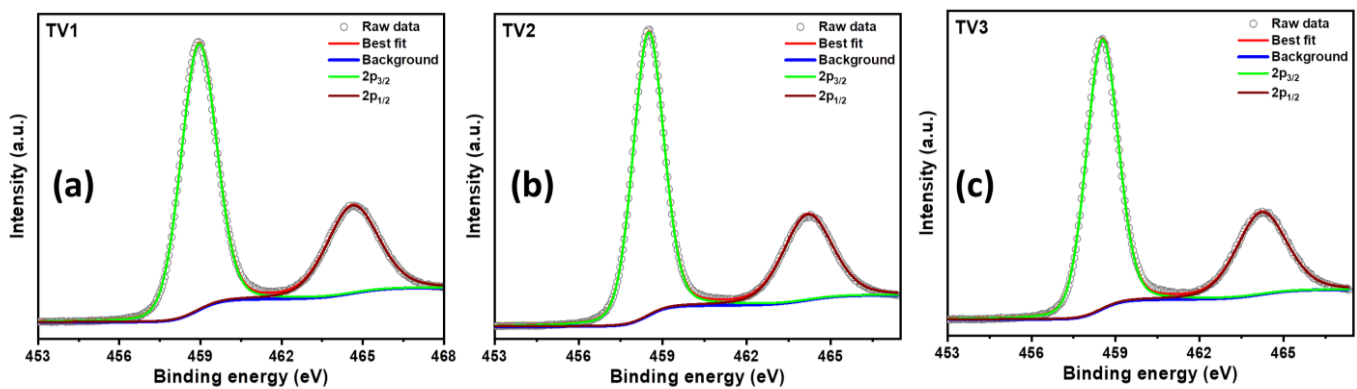

*Figure SM6: De-convoluted high resolution XPS spectra of Ti 2p for sample (a) TV1 (b) TV2 (c) TV3.*

*Table SM5: Fitting parameters for XPS spectra of Ti-2p*

| Samples | Peak 1 ($Ti^{4+}$ $2p_{3/2}$) | | | | Peak 2 ($Ti^{4+}$ $2p_{1/2}$) | | | | $\chi^2$ |
|---|---|---|---|---|---|---|---|---|---|
| | B.E. (eV) | FWHM (eV) | Area | G-L | B.E. (eV) | FWHM (eV) | Area | G-L | |
| TV0 | 458.93 | 1.2 | 120443.6 | 20 | 464.62 | 2.02 | 60221.88 | 28 | 10.884 |
| TV1 | 458.94 | 1.54 | 132949.1 | 15 | 464.63 | 2.25 | 66474.55 | 25 | 10.550 |
| TV2 | 458.49 | 1.3 | 186182.1 | 20 | 464.19 | 2.07 | 93091.05 | 28 | 14.068 |
| TV3 | 458.53 | 1.34 | 91255.05 | 17 | 464.22 | 2.11 | 45627.53 | 27 | 12.864 |
| TV4 | 458.49 | 1.35 | 167235 | 17 | 464.19 | 2.1 | 83617.51 | 28 | 21.693 |

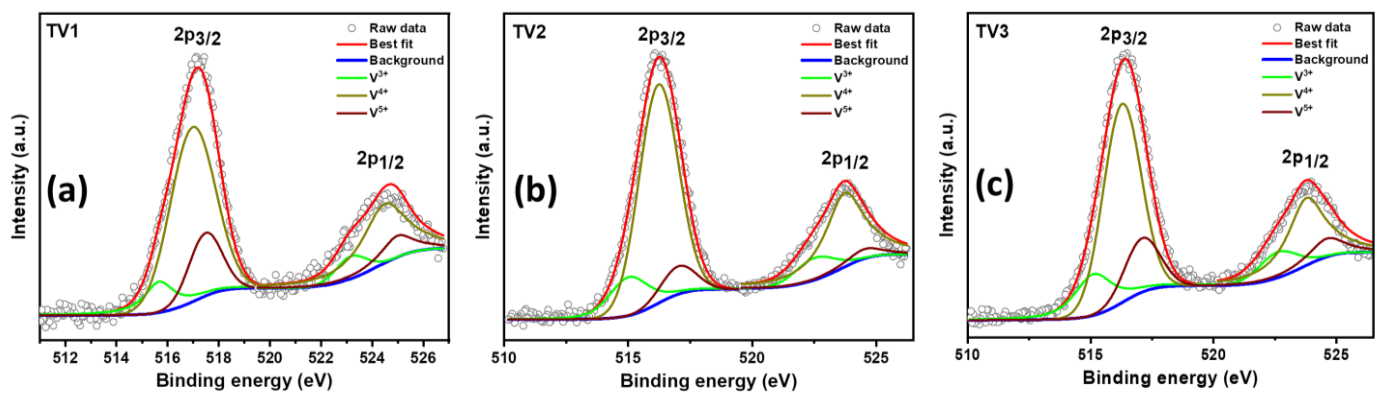

*Figure SM7: Fitted high-resolution XPS spectra showing the de-convoluted peaks and their corresponding chemical states for sample (a) TV1 (b) TV2 (c) TV3.*

*Table SM6: Fitting parameters for XPS spectra of V-2p*

| Samples | Peak 1 ($V^{3+}$ $2p_{3/2}$) | | | | Peak 2 ($V^{4+}$ $2p_{3/2}$) | | | | Peak 3 ($V^{5+}$ $2p_{3/2}$) | | | | |
|---|---|---|---|---|---|---|---|---|---|---|---|---|---|
| | B.E. (eV) | FWHM (eV) | Area | G-L | B.E. (eV) | FWHM (eV) | Area | G-L | B.E. (eV) | FWHM (eV) | Area | G-L | |
| TV1 | 515.67 | 1.41 | 1340.855 | 96 | 516.98 | 2.07 | 7650.708 | 0 | 517.48 | 1.51 | 2019.205 | 0 | |
| TV2 | 515.03 | 1.86 | 3480.133 | 31 | 516.22 | 1.87 | 17287.13 | 0 | 517.03 | 1.64 | 1944.286 | 0 | |
| TV3 | 515.12 | 1.83 | 2722.202 | 61 | 516.27 | 1.81 | 10050.65 | 0 | 517.10 | 1.68 | 2534.562 | 0 | |
| TV4 | 515.19 | 1.79 | 5078.175 | 55 | 516.28 | 1.82 | 28948.33 | 0 | 517.21 | 1.54 | 7157.564 | 0 | |
| Samples | Peak 4 ($V^{3+}$ $2p_{1/2}$) | | | | Peak 5 ($V^{4+}$ $2p_{1/2}$) | | | | Peak 6 ($V^{5+}$ $2p_{1/2}$) | | | | $\chi^2$ |
| | B.E. (eV) | FWHM (eV) | Area | G-L | B.E. (eV) | FWHM (eV) | Area | G-L | B.E. (eV) | FWHM (eV) | Area | G-L | |
| TV1 | 523.17 | 1.41 | 670.4273 | 0 | 524.48 | 2.07 | 3825.354 | 100 | 524.98 | 1.51 | 1009.602 | 100 | 2.544 |
| TV2 | 522.53 | 1.86 | 1740.067 | 0 | 523.72 | 1.87 | 8643.563 | 100 | 524.53 | 1.64 | 972.1431 | 100 | 3.912 |
| TV3 | 522.62 | 1.83 | 1361.101 | 8 | 523.77 | 1.81 | 5025.325 | 100 | 524.60 | 1.68 | 1267.281 | 100 | 3.657 |
| TV4 | 522.69 | 1.79 | 2539.088 | 0 | 523.78 | 1.82 | 14474.16 | 100 | 524.71 | 1.54 | 3578.782 | 100 | 6.065 |

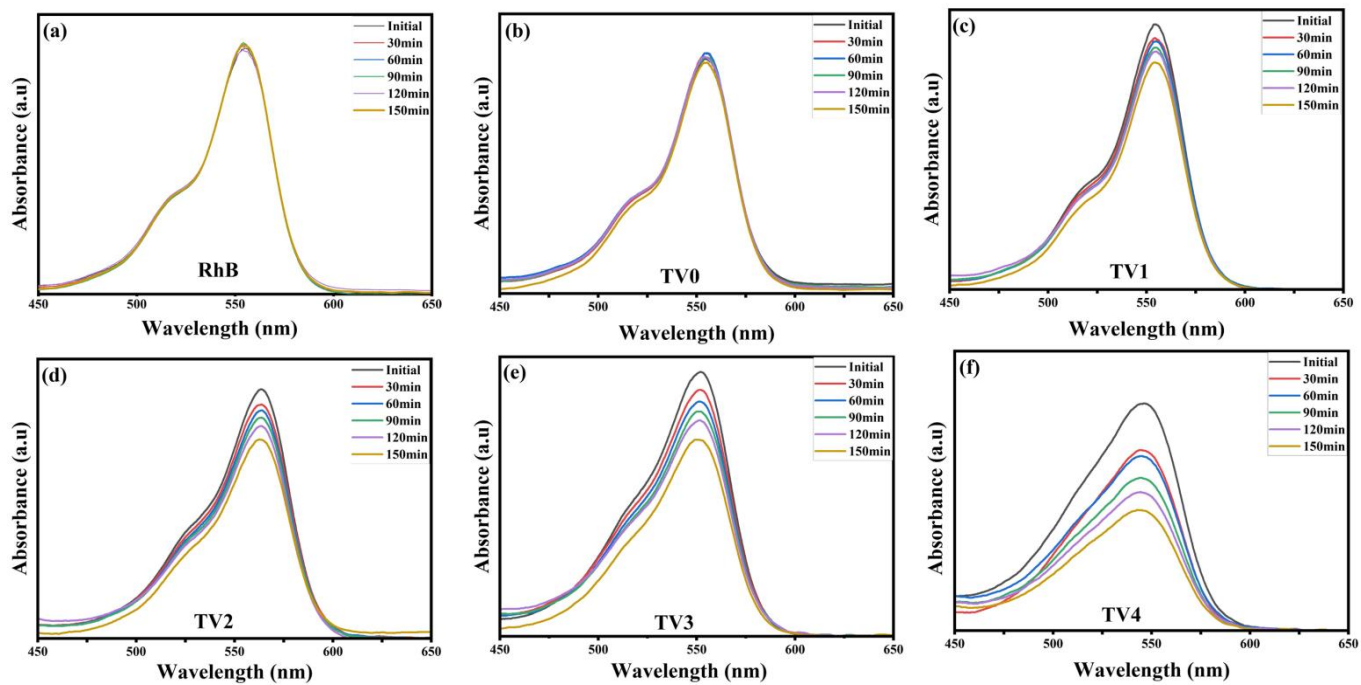

*Figure SM8: Recorded time evolution of the absorption spectra of RhB dye during the degradation process using different catalyst systems: (a) $H_2O_2$+No catalyst, (b) $H_2O_2$+TV0, (c) $H_2O_2$+TV1, (d) $H_2O_2$+TV2, (e) $H_2O_2$+TV3, and (f) $H_2O_2$+TV4.*

*Table SM7: Calculated dye-degradation percentages and apparent rate constants calculated based on $1^{st}$ order reaction kinetics.*

| Catalyst+$H_2O_2$ | Degradation ($\eta$) % in 150 min | Rate Constant ($k$) ($10^{-4}$ $min^{-1}$) |
|---|---|---|
| No Catalyst | 3.80 | 7.80 |
| TV0 | 7.60 | 5.37 |
| TV1 | 14.45 | 9.10 |
| TV2 | 23.92 | 17.90 |
| TV3 | 25.87 | 18.41 |
| TV4 | 51.20 | 96.90 |

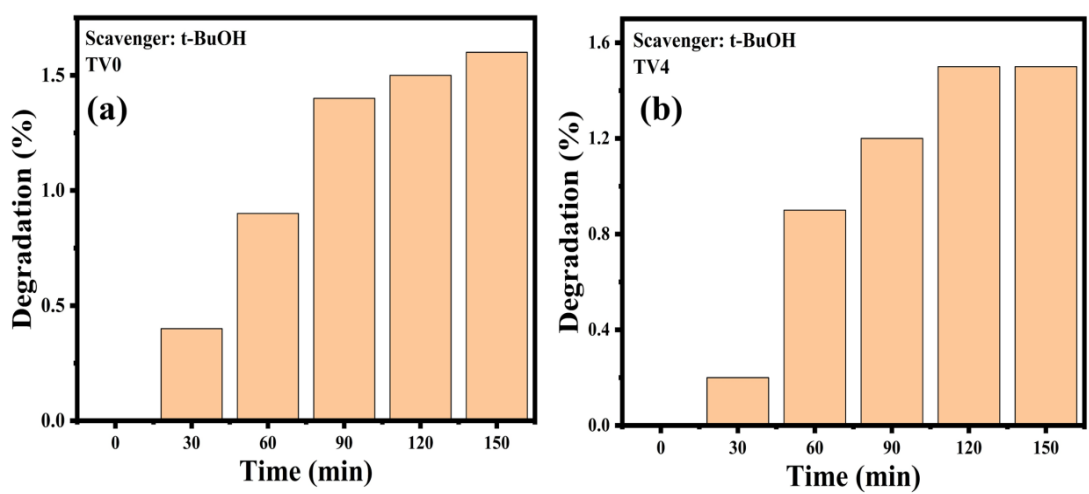

*Figure SM9: Time evolution of the degradation percentage in the presence of the •OH radical scavenger t-BuOH with catalysts (a) TV0, and (b) TV4.*

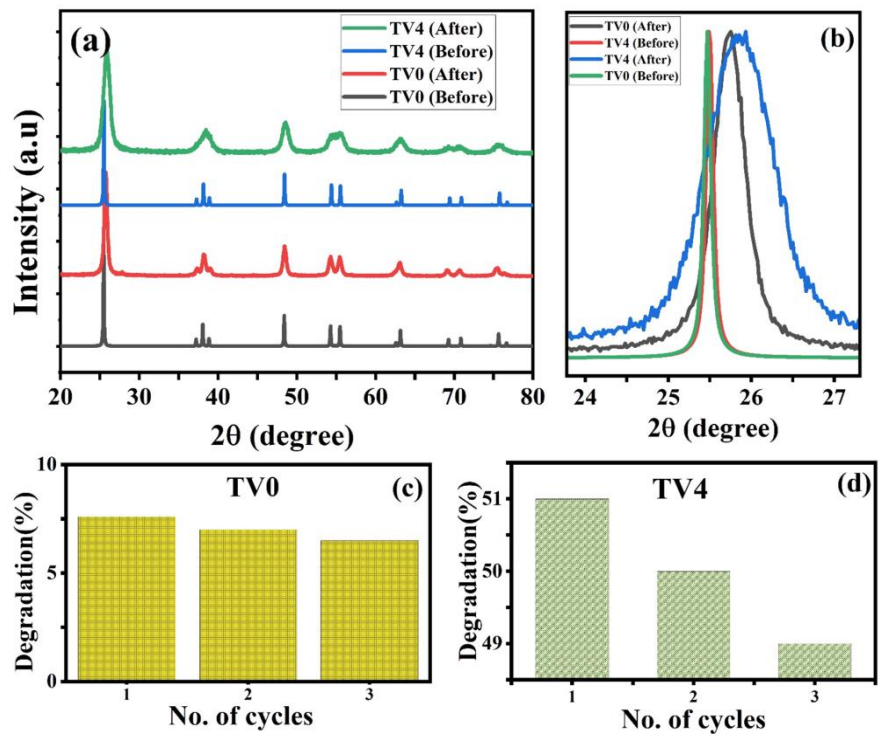

*Figure SM10: (a) XRD patterns of TV0 and TV4 before and after catalytic activity. (b) Broadening of highly intense (101) peak signifies a decrease in crystallinity after catalytic use. Negligible decrement in degradation efficiency over three consecutive cycles confirms the reusability of catalysts: (c) TV0 and (d) TV4.*